\documentclass[usenatbib]{mn2e_letter}
\usepackage[pdftex]{graphicx}
\usepackage{fixltx2e}
\usepackage{url}
\usepackage{amsmath,mathtools,revsymb,amssymb}
\usepackage{booktabs,multirow,calc,xspace,rotating}
\usepackage{microtype}

\usepackage{etoolbox}
\makeatletter

\patchcmd{\NAT@citex}
  {\@citea\NAT@hyper@{\NAT@nmfmt{\NAT@nm}\NAT@date}}
  {\@citea\NAT@nmfmt{\NAT@nm}\NAT@hyper@{\NAT@date}}
  {}
  {}

\patchcmd{\NAT@citex}
  {\@citea\NAT@hyper@{%
     \NAT@nmfmt{\NAT@nm}%
     \hyper@natlinkbreak{\NAT@aysep\NAT@spacechar}{\@citeb\@extra@b@citeb}%
     \NAT@date}}
  {\@citea\NAT@nmfmt{\NAT@nm}%
   \NAT@aysep\NAT@spacechar%
   \NAT@hyper@{\NAT@date}}
  {}
  {}

\patchcmd{\NAT@citex}
  {\@citea\NAT@hyper@{%
     \NAT@nmfmt{\NAT@nm}%
     \hyper@natlinkbreak{\NAT@spacechar\NAT@@open\if*#1*\else#1\NAT@spacechar\fi}%
       {\@citeb\@extra@b@citeb}%
     \NAT@date}}
  {\@citea\NAT@nmfmt{\NAT@nm}%
   \NAT@spacechar\NAT@@open\if*#1*\else#1\NAT@spacechar\fi%
   \NAT@hyper@{\NAT@date}}
  {}
  {}

\makeatother

\usepackage[dvipsnames]{xcolor}
\definecolor{halfgray}{gray}{0.55}
\definecolor{webgreen}{rgb}{0,.5,0}
\definecolor{webbrown}{rgb}{.6,0,0}
\usepackage[pdftex,hyperfootnotes=false,pdfpagelabels]{hyperref}
\hypersetup{%
    colorlinks=true, pdfstartview=FitV,%
    breaklinks=true, pdfpagemode=UseNone, pageanchor=true, pdfpagemode=UseOutlines,%
    plainpages=false,%
    bookmarksnumbered, bookmarksopen=true, bookmarksopenlevel=1,%
    hypertexnames=true, pdfhighlight=/O,%
    urlcolor=webbrown, linkcolor=RoyalBlue, citecolor=webgreen,%
    pdftitle={Thermal Instability},%
    pdfauthor={\textcopyright\ authors},%
    pdfsubject={},%
    pdfkeywords={},%
    pdfcreator={pdfLaTeX},%
    pdfproducer={LaTeX with hyperref}}

\newcommand{\ie}{i.\,e.\xspace}

\newcommand{\eg}{e.\,g.\xspace}

\newcommand{\cf}{cf.\xspace}

\newcommand{\fig}{Fig.\xspace}
\newcommand{\figs}{Figs.\xspace}

\newcommand\icm{\textsc{icm}\xspace}
\newcommand\rms{\textsc{rms}\xspace}
\newcommand\agn{\textsc{agn}\xspace}
\newcommand\mhd{\textsc{mhd}\xspace}
\newcommand\hbi{\textsc{hbi}\xspace}
\newcommand\mti{\textsc{mti}\xspace}
\newcommand\ti{\textsc{ti}}

\let\oldhat\hat
\renewcommand{\vec}[1]{\boldsymbol{#1}}
\renewcommand{\hat}[1]{\oldhat{\boldsymbol{#1}}}

\newcommand{\kperp}{\oldhat{k}_{\!\perp}}

\newcommand{\mr}[1]{\mathrm{#1}}

\newcommand{\ts}[1]{\ensuremath{t_{\mr{#1}}}}
\newcommand{\tsr}[2]{\ensuremath{\frac{\ts{#1}}{\ts{#2}}}}
\newcommand{\tsri}[2]{\ensuremath{\ts{#1}/\ts{#2}}}
\newcommand{\gr}[1]{\ensuremath{p_{\mr{#1}}}}

\newcommand{\scH}{\ensuremath{\mathcal{H}}\xspace}
\newcommand{\scL}{\ensuremath{\mathcal{L}}\xspace}

\newcommand{\unit}[1]{\ensuremath{\, \mathrm{#1}}}

\newcommand{\slfrac}[2]{\left. {#1} \middle/ {#2} \right.}

\newcommand{\ab}[1]{\langle {#1} \rangle}

\newcommand{\partiald}[2]{%
  \frac{\partial {#1}}{\partial {#2}}%
}
\newcommand{\partialD}[3]{%
  \left( \partiald{#1}{#2} \right)_{\!\!#3}%
}

\newcommand{\kb}{\ensuremath{k_{\mr{B}}}}
\newcommand{\mH}{\ensuremath{m_{\mr{H}}}}

\newcommand{\chie}{\ensuremath{\chi_{\mr{e}}}}

\newcommand{\Tfloor}{\ensuremath{T_{\mr{floor}}}\xspace}

\newcommand{\twod}{2\textsc{d}\xspace}
\newcommand{\threed}{3\textsc{d}\xspace}

\newcommand{\Qcond}{\ensuremath{\vec{Q}_{\mr{cond}}}\xspace}
\newcommand{\metricT}{\ensuremath{\mathbfss{I}}\xspace}

\newcommand{\ocool}{\ensuremath{\omega_{\mr{cool}}}\xspace}
\newcommand{\ocond}{\ensuremath{\omega_{\chi}}\xspace}

\newcommand{\halpha}{\ensuremath{\mathrm{H}_{\alpha}}\xspace}

\newcommand{\acknowledgments}{\begin{small}
    \section*{Acknowledgments}\end{small}}

\title[Thermal Instability in Gravitationally-Stratified Plasmas]{%
  Thermal Instability in Gravitationally-Stratified Plasmas: Implications for
  Multi-Phase Structure in Clusters and Galaxy Halos}

\newcommand\altaffilmark[1]{\textsuperscript{#1}} 
\newcommand\altaffiltext[1]{\textsuperscript{#1}}

\author[McCourt, Sharma, Quataert and Parrish]{
  \parbox[t]{0.9\textwidth}{
    \raggedright
    Michael McCourt,\altaffilmark{1}\thanks{E-mail:mkmcc@astro.berkeley.edu}
    Prateek Sharma,\altaffilmark{1,2}
    Eliot Quataert\altaffilmark{1} \&
    Ian J. Parrish\altaffilmark{1}}
  \vspace*{6pt} \\
  \altaffiltext{1}{Department of Astronomy and Theoretical Astrophysics
    Center, University of California
    Berkeley, Berkeley, CA 94720} \\
  \altaffiltext{2}{Einstein Fellow;
    Present Address:
    Department of Physics,
    Indian Institute of Science,
    Bangalore 560012, India} \\
}

\date{Submitted to MNRAS, May 2011}

\defcitealias{Sharma2011}{Paper~II}
\defcitealias{Balbus1989}{BS89}

\begin{document}
\maketitle
\label{firstpage}
\begin{abstract}
  We study the interplay among cooling, heating, conduction, and magnetic
  fields in gravitationally-stratified plasmas using simplified,
  plane-parallel numerical simulations.  Since the physical heating mechanism
  remains uncertain in massive halos such as groups or clusters, we adopt a
  simple, phenomenological prescription which enforces global thermal
  equilibrium and prevents a cooling-flow.  The plasma remains susceptible to
  \textit{local} thermal instability, however, and cooling drives an inward
  flow of material.  For physically plausible heating mechanisms in clusters,
  the thermal stability of the plasma is independent of its convective
  stability.  We find that the ratio of the cooling timescale to the dynamical
  timescale $\tsri{cool}{ff}$ controls the non-linear evolution and saturation
  of the thermal instability: when $\tsri{cool}{ff} \lesssim 1$, the plasma
  develops extended multi-phase structure, whereas when $\tsri{cool}{ff}
  \gtrsim 1$ it does not.  (In a companion paper, we show that the criterion
  for thermal instability in a more realistic, spherical potential is somewhat
  less stringent, $\tsri{cool}{ff} \lesssim 10$.)  When thermal conduction is
  anisotropic with respect to the magnetic field, the criterion for
  multi-phase gas is essentially independent of the thermal conductivity of
  the plasma.  Our criterion for local thermal instability to produce
  multi-phase structure is an extension of the cold vs. hot accretion modes in
  galaxy formation that applies at all radii in hot halos, not just to the
  virial shock.  We show that this criterion is consistent with data on
  multi-phase gas in the \textsc{accept} sample of clusters; in addition, when
  $\tsri{cool}{ff} \gtrsim 1$, the net cooling rate to low temperatures and
  the mass flux to small radii are suppressed enough relative to models
  without heating to be qualitatively consistent with star formation rates and
  x-ray line emission in groups and clusters.
\end{abstract}


\section{Introduction}
\label{sec:introduction}
While the formation of dark matter halos can be understood via gravitational
interactions alone, the combined effects of cooling and gravity are essential
to galaxy formation \citep{Rees1977,Silk1977,White1978}.  This interplay
remains poorly understood, however, because the dense plasma in many
high-mass halos is predicted to cool and accrete far more rapidly than is
observed \citep{Peterson2006}.  As a result, theoretical models and numerical
simulations routinely over-predict the amount of cooling and star formation
in massive galaxies \citep[\eg][]{Saro2006}; this discrepancy is an example
of the well-known ``cooling-flow problem.''  Some studies avoid the
cooling-flow problem by focusing only on very hot halos
\citep[\eg][]{Sijacki2006} or by ``pre-heating'' the gas to very high
entropies \citep[\eg][]{Oh2003,McCarthy2004}; this solution cannot work in
general, however, because such hot systems are not representative of the
cluster population \citep{Cavagnolo2008}.  In particular, the central cooling
time in many clusters is shorter than the Hubble time.  Significant heating
(``feedback'') is required even at low redshift to suppress cooling in
high-mass halos \citep{Benson2003}, and thus to explain the observed cutoff
in the galaxy luminosity function \citep{Cole2001,Kochanek2001}.

Detailed x-ray observations of groups and clusters also highlight the need
for significant heating of the intracluster plasma.  Though these objects
contain large amounts of radiating plasma \citep{Fabian1994}, their x-ray
spectra indicate a paucity of material cooling below $\sim 1/3$
of the maximum temperature \citep{Peterson2006}.  This demonstrates that most
of the plasma radiates without actually cooling; \ie, an energy source heats
the plasma at a rate similar to its cooling rate.

Although heating dramatically suppresses cooling in groups and clusters,
there is clear evidence for \textit{some} cool gas in these systems.
Studying this cold material can provide an important window into the heating
mechanisms in groups and clusters and may help us understand how the balance
between heating and cooling is maintained.  The existence of a cold phase can
be inferred from star formation \citep{ODea2010}, but it has also been
directly imaged in a number of cases, revealing filamentary nebulae located
tens of kiloparsecs from the center of the potential
\citep[\eg][]{Fabian2008,McDonald2010,McDonald2011}.  Despite more than five
decades of study, the origin of these dramatic filaments has yet to be
conclusively established: they have been interpreted as the remnant from an
enormous, central explosion \citep{Lynds1963,Lynds1970}, mass dropout from a
cooling catastrophe \citep{Fabian1977,Cowie1980,Nulsen1986}, debris from a
high-speed merger of two gas-rich galaxies \citep{Holtzman1992}, or material
dredged from the central galaxy by rising bubbles inflated by its \agn
\citep{Fabian2003,Fabian2008}.  Studies have shown, however, that the cold
gas is highly correlated with short central cooling times in the hot
intracluster plasma (\eg
\citealt{Hu1985,Heckman1989,Cavagnolo2008,Rafferty2008}, clearly illusrated
in \citealt{Voit2008}), suggesting that its origin involves cooling of the
intracluster medium~(\icm).

In this paper, we investigate the possibility that the cold phase forms as a
consequence of \textit{local} thermal instability in a \textit{globally
  stable} atmosphere.  Though many authors \citep[\eg][]{Fabian1977,Nulsen1986}
have previously proposed that the filaments form via thermal instability,
this idea has typically been analyzed in the context of a cooling-flow
background.  Subsequent analytic and numerical studies
\citep[\eg][]{Malagoli1987,Balbus1988,Balbus1989,Hattori1990,Malagoli1990,Joung2011}
showed, however, that the linear thermal instability is ineffective at
amplifying perturbations in a cooling flow and concluded that it is unlikely
unlikely to produce the cool filaments seen in many clusters.  By contrast,
the thermal instability is not suppressed in a globally stable atmosphere
\citep{Defouw1970,Balbus1986}, which is now believed to be a better
approximation to the thermal state of the \icm.  Quantitatively studying the
thermal instability in this context has proven difficult because of the
cooling-flow problem: studies that include cooling and gravity generally find
that the plasma is globally thermally unstable, and that the entire cluster
core collapses monolithically.\footnote{One-dimensional models of the \icm
  with simplified heating prescriptions can be stable or quasi-stable with
  episodes of heating and cooling \citep[\eg][]{Guo2008,Ciotti2001}; however,
  creating a realistic, stable model in two or three dimensions is
  significantly more challenging.  Moreover, just as convection cannot be
  modeled in one dimension, the dynamics of cool, over-dense gas sinking
  through the hot atmosphere is absent in one-dimensional models.}

We avoid the cooling-flow problem in this paper using a new strategy.  Rather
than attempting ab initio calculations of heating in clusters, we start from
the observational fact that the \icm does not cool catastrophically.  We
therefore implement a phenomenological heating model that enforces
approximate thermal equilibrium when averaged over large scales.  We use this
model to study the formation of multi-phase structure and we compare our
results with archival data for groups and clusters.  We find that the thermal
stability of the plasma does not depend on its convective stability (see
section~\ref{subsec:convection}).  Instead, we find that the ratio of the
thermal instability timescale $\ts{\ti}$ to the dynamical (or ``free-fall'')
timescale $\ts{ff}$ governs the non-linear saturation of the local thermal
instability: the plasma develops extended, multi-phase structure only where
this ratio falls below a critical threshold (\S\,\!\ref{subsec:multiphase}).
This conclusion is not sensitive to significant perturbations about our
idealized feedback prescription (\S\,\!\ref{subsec:heating-sensitivity}) and
is unchanged even in the presence of very rapid thermal conduction
(\S\,\!\ref{subsec:complex-results}) (Though the threshold may depend
somewhat on the geometry and initial conditions of the system;
see section~\ref{sec:discussion}).

This paper is the first in a series; here we present our model of local
thermal instability and demonstrate its properties and implications using
linear theory and non-linear simulations.  The aim of this paper is to
develop an understanding of the essential physics of the problem and we
therefore study stratified plasmas using idealized, plane-parallel
calculations.  In our companion paper (\citealt{Sharma2011}; hereafter
\citetalias{Sharma2011}), we present more realistic calculations of groups
and clusters with spherical geometries and NFW halos.  In both papers, we
focus our analysis on the transition of material from the hot phase to the
cold phase; we are not yet able to quantitatively predict any precise
properties (such as sizes or luminosities) of the cold filaments produced via
thermal instability.  We discuss in section~\ref{sec:discussion} how our
results can nonetheless be tested observationally.

Because we put in by hand that hot halos are in approximate global thermal
equilibrium, our model provides no direct insight into how this balance is
maintained.  This is both a weakness and a strength of our current approach:
though our setup is necessarily phenomenological, our results are not tied to
any particular heating mechanism.  Thus, we expect that our conclusions
should apply to a wide range of systems, ranging in mass from galaxies to
galaxy clusters.  We return to this point in
sections~\ref{sec:simulation-results} and~\ref{sec:discussion} and we study
more physically motivated heating models in \citetalias{Sharma2011}.  Our
present aim is not to identify a plausible solution to the cooling-flow
problem, but rather to understand what implications a stabilizing heat source
has for the local thermal stability and dynamics of the \icm.

The structure of this paper is as follows.  In section~\ref{sec:method}, we
describe our model for the plasma, including our phenomenological heating
prescription.  We describe our numerical method in section~\ref{sec:setup},
linear theory results in section~\ref{sec:linear-results}, and our primary
numerical results in section~\ref{sec:simulation-results}.
Section~\ref{sec:saturation} provides a physical interpretation of the
numerical results.  For simplicity, we initially ignore magnetic fields and
thermal conduction in this paper; section~\ref{sec:complex-sims} shows
results including these effects.  Finally, in section~\ref{sec:discussion},
we speculate on the astrophysical implications of our model and compare our
results with observational data from the \textsc{accept} catalog
\citep{Cavagnolo2009}.

\section{Plasma Model}
\label{sec:method}
In this section, we describe our model for the cooling, heating and dynamics
of the plasma in a dark matter halo.  Due to the wealth of observations of
the \icm, we explicitly motivate our model for galaxy clusters, and some of
the details we present in this section may not apply to galaxies.
Nonetheless, our analysis is fairly general and we expect that some of our
basic conclusions also hold massive galaxies (see \citetalias{Sharma2011} for
more details).

We model the plasma as an ideal gas, sitting in the fixed gravitational
potential of the halo and subject to both optically-thin radiative cooling
and heating by a stabilizing feedback mechanism.  In the interest of
simplicity, we initially ignore both thermal conduction and the dynamical
effect of the magnetic field; these effects are important in the \icm (see,
\eg \citealt{McCourt2011} and references therein), but do not change our
qualitative conclusions.  We generalize our results to conducting, magnetized
plasmas in section~\ref{sec:complex-sims}.

The equations for the conservation of mass and momentum in the plasma, and
for the evolution of its internal energy are:%
\begin{subequations}%
  \begin{align}%
    \partiald{\rho}{t} + \nabla\cdot(\rho\,\vec{v})
    &= 0\label{eq:consmass}, \\
    %
    \partiald{}{t} \left(\rho\,\vec{v}\right)
    + \nabla \cdot
    \left(\rho\,\vec{v}\otimes\vec{v} + P\,\metricT\right)
    &= \rho\,\vec{g},\label{eq:consmom} \\
    %
    \rho \, T \frac{d s}{d t}&
    = \scH-\scL\label{eq:inte},
  \end{align}\label{eq:dyn}
\end{subequations}
where $\rho$ is the mass density, $\vec{v}$ is the fluid velocity, $\otimes$
denotes a tensor product, $P$ is the pressure, \metricT is the unit matrix,
$\vec{g}$ is the gravitational field, $T$ is the temperature,
\begin{align}
  s = \frac{1}{\gamma-1} \frac{\kb}{\mu\,\!\mH}
  \ln\left(\frac{P}{\rho^{\gamma}}\right)\label{eq:entropy-defn}
\end{align}
is the entropy per unit mass, and $d/dt = \partial / \partial t +
\vec{v}\cdot\nabla$ is the Lagrangian (or convective) time derivative.  In
equation~\ref{eq:entropy-defn}, $\kb$ is Boltzmann's constant and
$\mu\,\!\mH$ is the mean mass of the particles contributing to thermal
pressure in the plasma.  The functions \scH and \scL describe heating and
cooling of the plasma, respectively; we explain our prescriptions for these
processes in the following sections.

\subsection{Feedback}
\label{subsec:feedback}
The physical origin of heating in clusters remains uncertain, but it is
simple to understand why our model requires the heating function \scH.
Equation~\ref{eq:inte} shows that the timescale for the \icm to cool in a
cluster is $\sim n T \left|\scL - \scH\right|^{-1}$; if $\scH = 0$, this
timescale near the centers of many clusters can be orders of magnitude
shorter than the Hubble time (this is the aforementioned cooling-flow
problem).  The continued existence of the \icm in these clusters therefore
strongly suggests that it is very nearly in thermal equilibrium, with $\scH$
approximately equal to $\scL$ when averaged over sufficient length- or
time-scales.\footnote{An alternative is a ``cooling-flow'' model, where the
  cooling gas flows inward and is replenished by continued accretion
  \citep[see][]{Fabian1994}.  This is not a viable alternative to heating,
  however, as these models over-predict the rate of gas cooling to low
  temperatures and the star formation rates in clusters by a factor of
  10--1000; furthermore, the resulting density profiles are strongly disfavored
  by x-ray observations \citep{McNamara2007}.  We therefore do not consider
  cooling-flow models in this work.}  Nonetheless, the multi-phase structure
seen in many clusters \citep[\eg][]{McDonald2010,McDonald2011} suggests that
the thermal instability also operates.  For the purposes of this paper, we
call this behavior globally stable, but locally thermally unstable.

The processes maintaining global thermal stability in clusters are not fully
understood.  The condition of global stability with local instability
constrains the possible heating mechanisms, however, and suggests a
phenomenological model for heating in the \icm.  This is a model in which
$\scH \approx \scL$ on average, but not $\scH = \scL$ identically.  We adopt
a specific implementation of this feedback model which simply fixes thermal
equilibrium at all radii in our model halos.  We set
\begin{align}
  \scH = \ab{\scL},\label{eq:simple-heating-fn}
\end{align}
where $\ab{\cdots}$ denotes a spatial average at a given radius.  Thus,
heating in our simplified model is a function only of $r$ and $t$.  By
construction, this heating function ensures global thermal equilibrium at all
radii in the plasma (precluding a cooling catastrophe), but permits the
thermal instability to grow on smaller scales.  It thus captures what we
believe is the essential physics for the formation of multi-phase structure
and meets our observationally-motivated requirements for the thermal
stability of the \icm.

Equation~\ref{eq:simple-heating-fn} can be roughly motivated by positing a
causal relationship between cooling on small scales and heating on large
scales.  Accretion onto a central \agn induced by cooling at larger radii is
a promising mechanism for this ``feedback''
\citep{Pizzolato2005,Pizzolato2010}, and feedback from star formation could
play a similar role in lower mass halos.  Our specific heating implementation
instantaneously balances cooling in every radial shell---in detail, this
behaviour is non-local, acausal, and unphysical.
Equation~\ref{eq:simple-heating-fn} is intended to mimic the end result of
very effective feedback, but does not directly model the feedback process.
Finding a physically-motivated heating mechanism that also leads to global
stability is an important goal in the theory of the \icm, but it is outside
the scope of our present study.

Our heating model is necessarily idealized, and it is important to separate
tautological results (put in by hand) from the results which more generally
reflect the global stability and local instability of the plasma.  Though the
subtleties of feedback are likely to strongly affect the evolution of the
plasma, we find that our qualitative conclusions are not sensitive to the
precise form of our heating function.  We demonstrate this in
section~\ref{subsec:heating-sensitivity} by applying spatial and temporal
variations to equation~\ref{eq:simple-heating-fn}.  Moreover, the simulations
in \citetalias{Sharma2011} reach similar conclusions using a very different
setup.  Thus we believe that the results derived using
equation~\ref{eq:simple-heating-fn} capture some of the essential (and
robust) dynamics of local thermal instability in globally stable systems.  It
is, however, difficult to prove this conclusively given current uncertainties
in the heating of the \icm.

In our heating model, spatial variations between heating and cooling drive
thermal instability; a more realistic model would likely introduce temporal,
in addition to spatial, variations.  We show
in~\S\,\!\ref{subsec:heating-sensitivity} that our conclusions do not change
unless these temporal fluctuations around the thermal equilibrium are very
large ($\sim300\%$).  We choose to begin our study using
equation~\ref{eq:simple-heating-fn} because it is analytically tractable and
lends itself to a thorough investigation.

Equation~\ref{eq:simple-heating-fn} is appropriate for a heating process
which distributes energy per unit volume, such as the dissipation of \mhd
waves.  Other processes like photoelectric heating distribute energy per unit
mass.  Since it is not yet known how feedback energy is thermalized in the
\icm, we generalize equation~\ref{eq:simple-heating-fn} to other processes:
\begin{align}
  \scH = n^{\alpha} \frac{\ab{\scL}}{\ab{n^{\alpha}}} .\label{eq:heating-fn}
\end{align}
Here, $\alpha = 0$ corresponds to volumetric heating and $\alpha = 1$
corresponds to mass-weighted heating.  We show in
section~\ref{subsec:stability} that the thermal instability takes $\sim3$
times longer to develop in plasmas with $\alpha = 1$ than in plasmas with
$\alpha = 0$; after scaling the timescales by this factor, however, we find
very similar evolution for plasmas with volumetric and mass-weighted heating
(see \fig~\ref{fig:simulation-snapshots}, below).

\subsection{Cooling}
\label{subsec:cooling}
In the idealized spirit of this paper, we adopt a simple cooling function
\scL dominated by thermal Bremsstrahlung
\begin{align}
  \scL_{\mr{B}} = n^2 \Lambda(T)
  = \Lambda_0 n^2 T^{1/2} ,\label{eq:raw-cooling-fn}
\end{align}
where $n = \rho / \mu\,\!\mH$ is the number density of particles in the
plasma and we have introduced the standard notation $\Lambda(T)$ for
consistency with other work.  Our conclusions are not sensitive to the shape
of the cooling function as long as the plasma remains locally thermally
unstable (\S\,\!\ref{sec:linear-results}); this is the case in the \icm for
temperatures above $\sim 10^4\unit{K}$.

In an unstratified plasma, thermally unstable clumps of cool gas collapse to
the Field length in the cold phase (the length-scale below which thermal
conduction suppresses local thermal instability; \citealt{Field1965}).
Resolving the realistic Field length in the cold phase of the \icm is
numerically impractical, so we introduce a temperature floor at which we
truncate the cooling function \citep[see][\S\,\!2.2 for a discussion of this
approximation; also see \S\,\!\ref{subsec:resolution-study} of this
paper]{Sharma2010}.  We use the modified cooling function
\begin{align}
  \scL = \scL_{\mr{B}} \,\Theta_{\mr{H}}(T-\Tfloor) ,\label{eq:cooling-fn}
\end{align}
where $\Theta_{\mr{H}}$ is the Heaviside function, and \Tfloor effectively
becomes the temperature of the cold phase.

The microphysical processes heating and cooling the cold phase in the \icm
are likely to be very complicated \citep[see][]{Ferland2009} and we do not
consider them here.  Our use of a temperature floor amounts to the reasonable
assumption that, once a thermally-unstable fluid element cools below
$\Tfloor$, it is unlikely to enter back into the hot phase.  This
simplification, along with our omission of line-cooling from
equation~\ref{eq:raw-cooling-fn}, prevents us from studying the evolution of
the cold material in detail.  This is not a major limitation, however,
because we are primarily interested in the transition of material from the
hot phase to the cold phase.  Following the internal structure of the cold
clumps would be crucial for calculating the emission from filaments or for
studying the intermittency in the accreted mass flux, but these applications
are beyond the scope of our present study.

The simplified cooling function used here (eq.~\ref{eq:cooling-fn}) prevents
the gas from cooling below \Tfloor and therefore artificially lowers the gas
density in thermally unstable clumps or filaments.  We have confirmed,
however, that the quantitative results in this paper are insensitive to the
numeric value of \Tfloor, provided it is much lower than the initial (or
virial) temperature of the plasma.  In \citetalias{Sharma2011} we use a
realistic cooling function that includes both Bremsstrahlung and line
emission, and which does not implement a temperature floor.  The results from
this more realistic model agree with our conclusions here.

\section{Numerical Model}
\label{sec:setup}
We solve equations~\ref{eq:consmass}--\ref{eq:inte} using the conservative
\mhd code \textsc{Athena}, modified to implement
equations~\ref{eq:heating-fn} and~\ref{eq:cooling-fn} via a semi-implicit,
operator-split method \citep{Sharma2010}.  Specifically, we evolve the
thermal energy per unit volume $E = n \, \kb T / (\gamma-1)$ using
\begin{subequations}
  \begin{align}
    \delta E^{(n)} &= \left( \scH^{(n)} - \scL^{(n)} \right) \, \delta t \\
    E^{(n+1)} &=
    \begin{dcases}
      E^{(n)} + \delta E^{(n)}
      & \delta E^{(n)} > 0 \\
      E^{(n)} / \left( 1 + |\delta E^{(n)} / E^{(n)}| \right)
      & \delta E^{(n)} < 0
    \end{dcases} ,
  \end{align} \label{eq:semi-implicit-method}
\end{subequations}
where $\delta$ indicates a finite approximation to a differential, and
$f^{(n)}$ denotes the function $f$ during the $n$\textsuperscript{th}
time-step of the simulation.  This method explicitly prevents the temperature
from becoming negative, even in the extreme case that the cooling time
becomes shorter than the simulation time-step (although
equation~\ref{eq:semi-implicit-method} is no longer accurate in this limit).
Equation~\ref{eq:semi-implicit-method} is asymmetric and is only accurate to
first order in $\delta t / \ts{cool}$.  In order to test the sensitivity of
our simulations to these shortcomings, we have also run simulations using a
fully explicit, sub-cycled method.  The two methods yield very similar
results.  We use equation~\ref{eq:semi-implicit-method} because it is faster
than an explicit method and because it does not alter our results.

We perform most of our calculations on \twod Cartesian grids of resolution
$(300)^2$ or \threed Cartesian grids of resolution $(128)^3$.  We show a
resolution study in section~\ref{subsec:resolution-study}.  In the remaining
sections, as in our simulations, we work in units with $k_{\mr{B}} =
\mu\,\!m_{\mr{p}} = 1$.

We perform our calculations in the plane-parallel approximation, with
$\vec{g} = - g(z) \, \hat{z}$.  We therefore use the words `height' and
`radius' interchangeably in the following sections.  We make our setup
symmetric about the $z = 0$ plane, with
\begin{align}
  g = g_0 \frac{z / a}{\left[1 + (z/a)^2\right]^{1/2}} .\label{eq:gravity-defn}
\end{align}
Thus, $g$ is nearly constant outside $|z| = a$, with a smooth transition
through zero at the center.  This setup enables us to place the computational
boundaries far from the center, where most of the cooling and feedback take
place (see \fig~\ref{fig:simulation-snapshots-detail}).  To further diminish
the influence of the boundaries, we end our simulations before one cooling
time transpires at the boundary.  We use reflecting boundary conditions in
the direction parallel to gravity and periodic boundary conditions in the
orthogonal directions.

We set the softening radius $a = 0.1 \, H$, where $H$ is the plasma
scale-height (defined below).  We turn off cooling and heating within $|z|
\leq a$ because the physics at small radii is particularly uncertain and our
feedback prescription (equation~\ref{eq:heating-fn}) may not be a good
approximation to what happens there.  We allow cold material to accumulate in
the center $|z| \leq a$, but we otherwise ignore this region in our analysis.
We have also performed simulations in which we do not turn off cooling in the
center and have confirmed that it does not change our conclusions at larger
radii $z \gtrsim H$.

We initialize the \icm in hydrostatic equilibrium, with a constant temperature
$T_0$ and with the density profile
\begin{align}
  \rho(z) = \rho_0 \, \exp\left[ -\frac{a}{H}
    \left(\left[1 + (z/a)^2\right]^{1/2}-1\right)\right] ,\label{eq:isothermal-atm}
\end{align}
where the scale-height $H = T_0/g_0$.  For computational convenience, we set
$\rho_0 = T_0 = g_0 = 1$ and we take $\Tfloor = 1/20$.  This roughly
corresponds to a virialized halo, in which the thermal and gravitational
energy in the plasma are approximately equal.  The atmosphere defined by
equation~\ref{eq:isothermal-atm} is buoyantly stable, with $\partial s /
\partial z > 0$.  To test the sensitivity of our results to stratification,
we also use the buoyantly neutral atmosphere defined by
\begin{subequations}
  \begin{align}
    T(z) &= T_0 \left[ 1 - \frac{\gamma-1}{\gamma} \frac{a}{H}
      \left(\left[1 + (z/a)^2\right]^{1/2}-1\right) \right], \\
    \rho(z) &= \rho_0 \left( \frac{T}{T_0} \right)^{\slfrac{1}{(\gamma-1)}} .
  \end{align}\label{eq:isentropic-atm}
\end{subequations}
We refer to the conditions defined by equations~\ref{eq:isothermal-atm}
and~\ref{eq:isentropic-atm} as isothermal and isentropic, respectively.  Note
that our use of the entropy gradient to determine convective stability is
only appropriate because we have neglected conduction.  When thermal
conduction is efficient, the temperature gradient and the magnetic field
orientation determine the convective stability of the plasma
\citep{Balbus2000,Quataert2008}.  We describe this in more detail in
section~\ref{sec:complex-sims}.

We seed thermal instability in our model atmospheres by applying an isobaric
perturbation with a flat spectrum ranging from $k = 2\pi/L$ to $k = 40\pi/L$,
where $L$ is the size of the simulation domain.  The cutoff at high $k$ makes
the perturbation independent of resolution and permits a detailed convergence
study.  Unless otherwise noted, the modes of this perturbation have
Gaussian-random amplitudes with an \rms value of~$10^{-2}$.

We define the free-fall time and the cooling time as follows:
\begin{subequations}
  \begin{align}
    \ts{ff}&   = \left(\frac{2 z}{g_0} \right)^{1/2} \\
    \ts{cool}& = \frac{3}{2} \frac{T^{1/2}}{n \Lambda_0} .
  \end{align}
\end{subequations}
Since these timescales are functions of height in our simulations, we quote
them in the plane $z = H$ to give single values.  When our analysis depends
on the ratio $\tsri{\ti}{ff}$, we restrict it to this plane.  We perform
simulations with different initial values of the ratio $\tsri{cool}{ff}$ by
changing the parameter $\Lambda_0$; this permits direct and unambiguous
comparison among our simulations because each is initialized identically.
In reality, of course, $\Lambda_0$ is set by fundamental physics, and
different values of $\tsri{cool}{ff}$ correspond to clusters with different
\icm entropies or densities.

\section{Linear Theory Results}
\label{sec:linear-results}
Equations~\ref{eq:dyn}--\ref{eq:cooling-fn} completely specify our model.  In
the rest of this paper, we study the properties of this model and apply it to
astrophysical systems.  In this section, we describe the linear stability of
our model and derive the timescale for the formation of multi-phase structure
in the plasma.  We discuss the well-known linear results in some detail
because they inform our interpretation of the non-linear behavior described
later.  In addition, the interpretation of these linear results has generated
some confusion in the literature, leading to conflicting claims about the
thermal stability of gas in hot halos.

\subsection{Linear Stability}
\label{subsec:stability}
We define the net cooling rate $\Theta = \scL - \scH$ and assume that the
plasma is initially in thermal equilibrium with $\Theta = 0$ everywhere.  The
derivative $(\partial \Theta / \partial T)_P$ describes how the net cooling
responds to a linear, Eulerian perturbation.  If this derivative is negative,
a decrease in temperature at a fixed location in the plasma leads to an
increase in the net cooling rate; thus, the temperature decreases further and
the perturbed fluid element runs away to low temperatures.  Similarly, an
increase in temperature causes the fluid element to run away to high
temperatures.  The plasma is therefore unstable to small temperature
fluctuations when $(\partial \Theta / \partial T)_P < 0$.  A similar line of
reasoning demonstrates that the plasma is thermally stable if $(\partial
\Theta / \partial T)_P > 0$.  Following \citet{Field1965}, we derive this
result by linearizing and perturbing
equations~\ref{eq:consmass}--\ref{eq:inte}.  This analysis yields the linear
growth rate of the perturbations and will assist our interpretation of the
non-linear results presented later.

We Fourier transform equations~\ref{eq:consmass}--\ref{eq:inte} and perform a
standard WKB analysis.  We seek solutions with growth times much longer than
the sound-crossing time and therefore make the Boussinesq approximation,
which filters out sound waves \citep{Balbus2000,Balbus2001}.  Under these
approximations, the dynamical equations become
\begin{subequations}
  \begin{align}
    \vec{k} \cdot \delta \vec{v} &= 0 \label{eq:pert-mass} \\
    -i \, \omega k^2 \delta \vec{v} &= -\frac{\delta n}{n}
    \left[ k^2 \vec{g} - \vec{k} (\vec{k} \cdot \vec{g}) \right]
    \label{eq:pert-mom} \\
    -i \, \omega \delta s + \delta v_z \partiald{s}{z} &=
    - \frac{\delta \Theta}{n T} .\label{eq:pert-e}
  \end{align}\label{eq:wkb}
\end{subequations}
In deriving equation~\ref{eq:pert-mom}, we have crossed the momentum equation
with $\vec{k}$ twice and used equation~\ref{eq:pert-mass} to eliminate the
compressive component of the velocity.  This is consistent with the
Boussinesq approximation and simplifies the algebra later on.  Additionally,
in the Boussinesq limit,
\begin{align}
  \delta s =
  -\frac{\gamma}{\gamma-1} \frac{\delta n}{n} \label{eq:pert-s}
\end{align}
and
\begin{align}
  \delta \Theta =
  -T \partialD{\Theta}{T}{P} \frac{\delta n}{n} .\label{eq:pert-theta}
\end{align}
In deriving equation~\ref{eq:pert-theta}, we have used the thermodynamic
identity
\begin{align}
  \partialD{\,\ln X}{\,\ln T}{P} =
  \partialD{\,\ln X}{\,\ln T}{n} \!-
  \partialD{\,\ln X}{\,\ln n}{T}
\end{align}
for any state function $X(n,T)$.  Note that, although the net cooling rate
$\Theta$ varies explicitly with position, this dependence does not enter into
equation~\ref{eq:pert-theta} because $\delta \Theta$ represents an Eulerian
perturbation at a fixed point in space.  Although heating in our model has
explicit radial and temporal dependencies, it experiences no first-order
change under an Eulerian perturbation.  We therefore ignore changes to the
heating in this linear analysis.  This should not give the impression that
heating is immaterial to the linear results; on the contrary, these results
presume an initial equilibrium state with a stabilizing heat source.  The
growth of the thermal instability is very different in the absence of such
heating \citep{Balbus1988,Balbus1989}.

Combining equations~\ref{eq:wkb}--\ref{eq:pert-theta}, we find that the
linear dispersion relation for the plasma is
\begin{align}
  \omega^2 - i \, \frac{T}{\scL}\partialD{\Theta}{T}{P} \omega \ocool
  - N^2 \left( 1 - \oldhat{k}_z^2 \right) = 0 \label{eq:linear-dispersion}
\end{align}
where
\begin{align}
  \ocool = \frac{\gamma-1}{\gamma} \frac{\scL}{n T} = (\gamma\,\ts{cool})^{-1}
\end{align}
is the cooling rate,
\begin{align}
  N = \sqrt{ \frac{\gamma - 1}{\gamma} \; g \,
    \frac{\partial s}{\partial z} }\label{eq:brunt}
\end{align}
is the frequency for internal gravity waves, and $\hat{k} = \vec{k}/k$ is the
direction of the wave vector of the perturbation.  As noted previously, we
have neglected conduction and thus equations~\ref{eq:linear-dispersion}
and~\ref{eq:brunt} only apply on relatively large length scales, $\gtrsim$
the Field length (see section~\ref{sec:complex-sims}).
Equation~\ref{eq:linear-dispersion} implies that perturbations grow
exponentially in amplitude $\sim e^{\gr{\ti} t}$, with
\begin{subequations}
  \begin{align}
    \gr{\ti}  &= - \frac{\gamma-1}{\gamma} \frac{1}{n}
    \partialD{\Theta}{T}{P} \label{eq:field-growth-rate-super-general} \\
    &= \phantom{-} \frac{\gamma-1}{\gamma}
    \left( 2-\partiald{\, \ln \Lambda}{\, \ln T} - \alpha \right)
    \frac{\scL}{n T} \label{eq:field-growth-rate-general} \\
    &= \phantom{-} \left( \frac{3}{2} - \alpha \right)
    \ocool .\label{eq:field-growth-rate-specific}
  \end{align}\label{eq:field-growth-rate}
\end{subequations}
The three forms of equation~\ref{eq:field-growth-rate} are equivalent and
are useful in different contexts.  In
equation~\ref{eq:field-growth-rate-specific}, we have specialized to
Bremsstrahlung cooling.  In this case, plasmas with $\alpha < 3/2$ are
locally thermally unstable (with $\gr{\ti} > 0$), even though our model is
(by construction) globally stable against a cooling catastrophe.

\subsection{Local Stability, Global Stability, and Convection}
\label{subsec:convection}
Following \citet{Field1965} and \citet{Defouw1970}, we showed in the previous
section that the \icm is likely to be locally thermally unstable, and we
propose that thermal instability may produce at least some of the multi-phase
structure in galaxy clusters.  At first, our analysis may appear inconsistent
with other claims (such as can be found in, \eg \citealt{Balbus1989} and
\citealt{Binney2009}) about the importance of local thermal instability in
galaxy and cluster halos.  We review this apparent contradiction here and
show that there is no inconsistency.

\citet{Balbus1988} and \citet{Balbus1989} extensively studied thermal
instability using Lagrangian techniques and discovered that it is
significantly stabilized in a cooling-flow.  In a globally stable atmosphere,
however, perturbations \emph{do} grow exponentially
\citep{Defouw1970,Balbus1986}.  Since it is now thought that clusters are
globally thermally stable and that the \icm persists for many cooling times,
we expect the thermal instability to undergo many e-foldings and to become
highly non-linear in clusters (though this does not always imply a large
amplitude; see~\S\,\!\ref{sec:saturation}).  Thus, assumptions about the
global stability of the \icm also dictate conclusions about its local thermal
stability and one must be careful to choose an appropriate background model.

Even though we expect perturbations to grow exponentially in clusters, they
do not necessarily grow monotonically: equation~\ref{eq:linear-dispersion}
shows that a thermally unstable perturbation oscillates as it grows if the
cooling time is longer than the buoyancy time.  This overstability represents
a driven gravity wave \citep[see][]{Defouw1970}.  Since the thermal
instability in this case is not purely condensational, its identification
with multi-phase gas becomes somewhat unclear
\citep{Malagoli1987,Binney2009}.  However, the growth rate of the thermal
instability is essentially unaffected by buoyancy
(equation~\ref{eq:linear-dispersion}), and thus perturbations are also likely
to become highly non-linear in this limit.  Earlier studies of the thermal
overstability in stratified plasmas have either focused entirely on the
linear evolution of perturbations \citep{Defouw1970,Malagoli1987,Binney2009}
or have studied them in the context of a cooling-flow
\citep{Hattori1990,Malagoli1990,Joung2011}, in which the thermal instability
is suppressed \citep{Balbus1989}.

For the reasons listed above, we argue that earlier studies cannot directly
predict the astrophysical implications of thermal instability in cluster
halos.  The astrophysical implications of the thermal instability depend on
how the linear growth saturates in a globally stable environment.  This
motivates our present study.  A series of previous investigations are very
similar to ours \citep{Nulsen1986,Pizzolato2005,Soker2006,Pizzolato2010}, but
focus on the survival of preexisting cold filaments rather than their
formation via thermal instability.  Our investigation compliments these
studies and produces the initial conditions they require.

The saturation of the thermal instability involves the sinking of cool
over-densities; in this respect, it bears some similarity to convection.
This connection between thermal and convective stability was first recognized
by \citet{Defouw1970} and was significantly sharpened by \citet{Balbus1989}.
Specifically, \citet{Balbus1989} showed that thermal instability necessarily
implies convective instability if the heating and cooling are state functions
of the plasma.  Heating in galaxy groups and clusters is very unlikely to be
a state function of the \icm plasma, however.  As a concrete example of
spatially dependent heating, consider heating by turbulence (induced by, \eg
buoyant bubbles created by star formation or an \agn).  The heating rate in
this case is set by the rate at which turbulent energy is transferred to
small scales, and thus by the turbulence properties as a function of
position.  In this case, \ie when the heating depends explicitly on position,
there is no one-to-one relationship between convective and thermal stability
(as noted by \citealt{Balbus1989}).  Fundamentally, buoyancy determines
convective stability, while heating and cooling determine thermal stability;
these processes are not related in a globally stable atmosphere, and the
thermal stability of an atmosphere is independent of its convective
stability.

\section{Simulation Results}
\label{sec:simulation-results}
%
\begin{table}
  \caption{Parameters for simulations without
    conduction~(\S\,\!\ref{sec:simulation-results}).}
  \label{tab:simple-sims}
  \begin{center}
    \begin{tabular}{@{}cccc@{}}
      \toprule
      \parbox{\widthof{Condition}}{\centering Initial \\ Condition}
      & $L/H$ & $\alpha$ & $\tsri{\ti}{ff}$ \\
      \midrule
      \multirow{2}*[0ex]{Isothermal} & \multirow{2}*[0ex]{3} & 0  & $1.57 \, \Lambda_0^{-1}$ \\
      \cmidrule(l){3-4}
      &                      & 1  & $4.70 \, \Lambda_0^{-1}$ \\
      \midrule
      \multirow{2}*[0ex]{Isentropic} & \multirow{2}*[0ex]{2} & 0  & $1.25 \, \Lambda_0^{-1}$ \\
      \cmidrule(l){3-4}
      &                      & 1  & $3.74 \, \Lambda_0^{-1}$ \\
      \bottomrule
    \end{tabular}
  \end{center}

  \medskip
  We performed all simulations on square Cartesian grids of resolution
  $(300)^2$ or $(128)^3$ and physical size $2 L$ (the scale-height $H$ is
  defined in \S\,\!\ref{sec:setup}).  We also performed simulations at other
  resolutions as part of a convergence study
  (\S\,\!\ref{subsec:resolution-study}).  The cooling constant $\Lambda_0$ is
  a free parameter in our model, which we choose to obtain the desired
  $\tsri{\ti}{ff}$.  Each combination of the listed parameters was simulated
  with initial values of $\log_{10}(\tsri{\ti}{ff})$ at $z = H$ spanning
  between $-1$ and $1$ with increments of $1/4$.  The top row
  represents our fiducial setup; we also performed \threed simulations using
  this setup with $\log_{10}(\tsri{\ti}{ff}) =$ $-1$, $-0.75$, $-0.5$, $0$,
  and $1$.
\end{table}
%
\begin{figure*}
  \centering
  \includegraphics[width=7.0in]{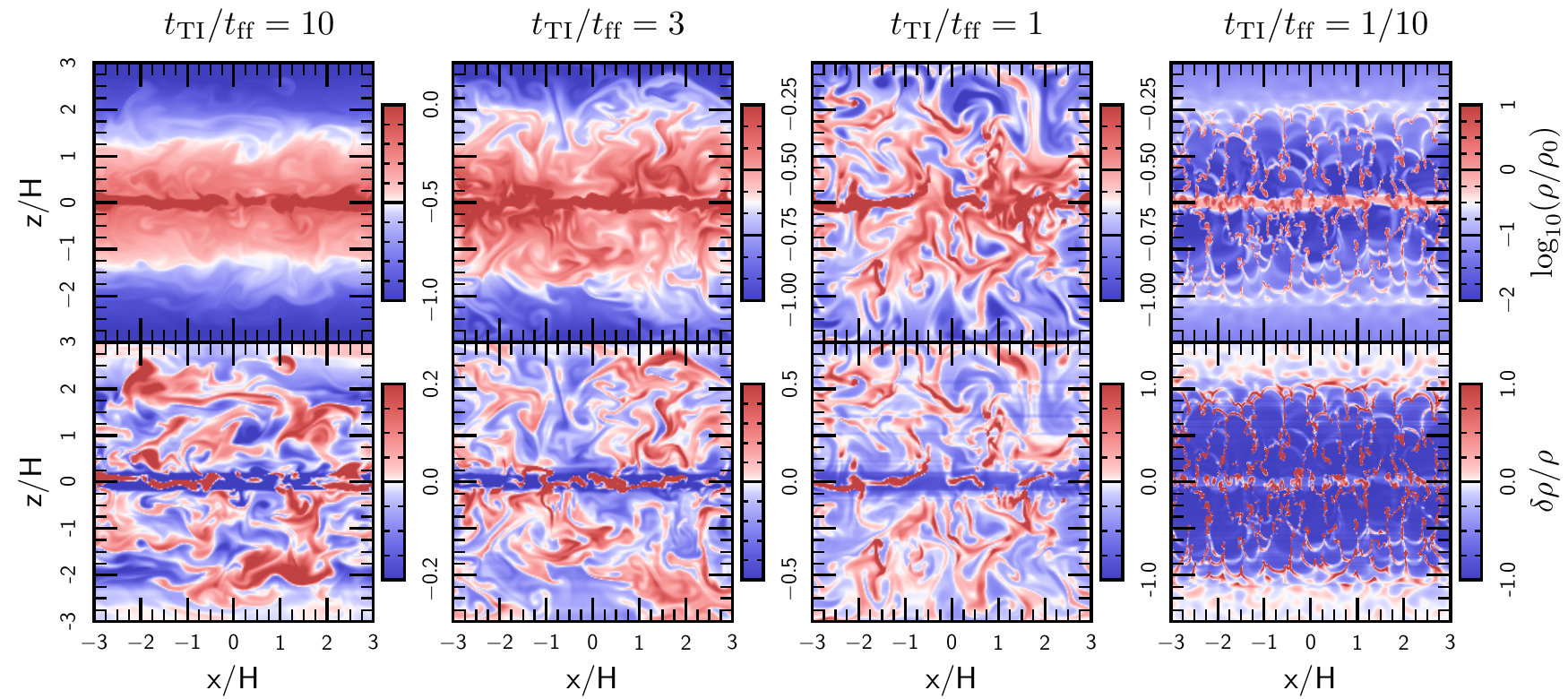}
  \caption{Snapshots of the density (top) and fractional density
    inhomogeneity $\delta \rho / \rho \equiv (\rho - \ab{\rho})/\ab{\rho}$
    (bottom) at the time $t = 10 \, \ts{\ti} (z = H)$ in our simulations.
    Note that our simulations are symmetric about the plane $z = 0$; this
    enables us to put the boundaries far from the center, where most of the
    cooling and feedback take place.  Gravity points down in the top half of
    the domain and up in the bottom half (for the remainder of the paper, we
    primarily show images of the top of the domain).  We applied the heating
    function in equation~\ref{eq:heating-fn} to ensure global thermal
    stability, distributing the energy per unit volume ($\alpha = 0$).  From
    left to right, these simulations have initial values of $\tsri{\ti}{ff}
    = 10, 3, 1$ and $0.1$.  These simulations demonstrate that cooling and
    heating drive internal gravity waves when $\tsri{\ti}{ff} \gtrsim 1$.
    The amplitude of these waves increases with the cooling rate and
    approaches the size of the simulation domain when $\ts{\ti} \sim
    \ts{ff}$.  When $\tsri{\ti}{ff} < 1$, the thermally unstable gas
    collapses into dense clumps, which then rain down into the center of the
    potential.  For clarity, we have restricted the color bar on plots with
    $\tsri{\ti}{ff} = 1/10$.  In this simulation, $(\delta \rho /
    \rho)_{\mr{max}} \sim 20$ is set by the (arbitrary) temperature floor we
    impose.  While the simulations have been run for 10 thermal instability
    times at $z = H$, gas near the boundaries has not yet had time to cool.
    The initial perturbations are still visible near the boundaries in the
    lower-rightmost plot.  This figure also clearly shows the accumulation
    of cool material in the center of our simulation domain.  We describe
    this process in more detail in~\S\,\!\ref{sec:saturation}.  Animated
    versions of the figures in this paper can be found at:
    \href{http://astro.berkeley.edu/~mkmcc/research/thermal\_instability/movies.html}%
    {http://astro.berkeley.edu/$\sim$mkmcc/research/thermal\_instability/movies.html}.}%
  \label{fig:simulation-snapshots-detail}
\end{figure*}
%
\begin{figure*}
  \centering
  \includegraphics[width=7.0in]{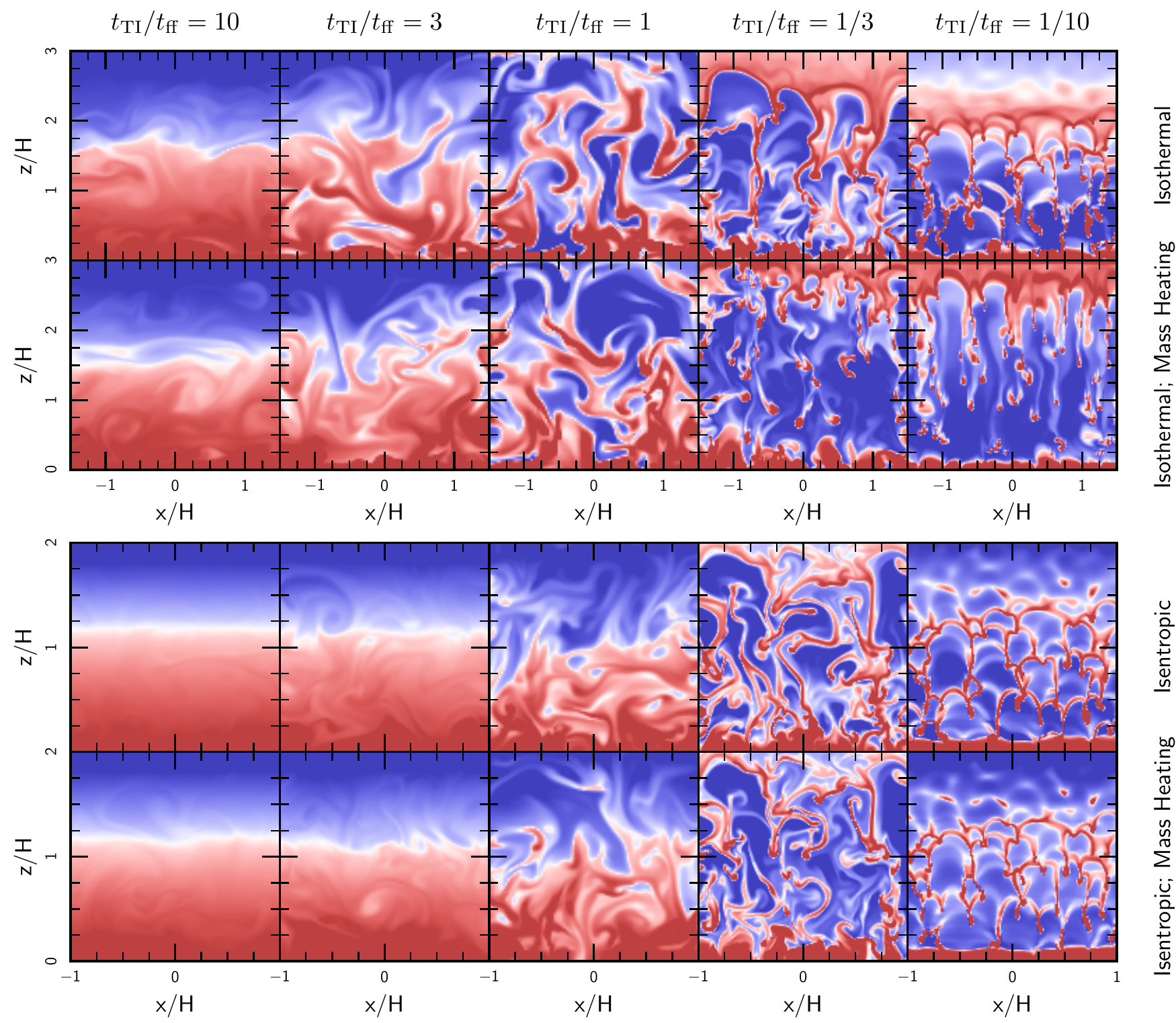}
  \caption{Snapshots of the density at the time $t = 10 \, \ts{\ti}$, for
    different values of the time-scale ratio $\tsri{\ti}{ff}$.  The top two
    rows show our simulations with isothermal initial conditions and volume
    and mass-weighted heating, while the bottom two rows show isentropic
    initial conditions with volume and mass-weighted heating.  These results
    show that the non-linear behavior of the thermal instability is
    relatively independent of the initial stratification and the details of
    the heating.  Note that the color scale was chosen to show the features
    in the gas and varies from plot to plot; the ranges for a given value of
    $\tsri{\ti}{ff}$ are similar to those shown in
    \fig~\ref{fig:simulation-snapshots-detail}. }%
  \label{fig:simulation-snapshots}
\end{figure*}
We extend our analysis into the non-linear regime using the numerical setup
described in section~\ref{sec:setup}.  We have run a large suite of \twod and
\threed simulations, summarized in Table~\ref{tab:simple-sims}.  We focus our
analysis on the presence of multi-phase
structure~(\S\,\!\ref{subsec:multiphase}) and on the accreted mass
flux~(\S\,\!\ref{subsec:mass-flux}), both of which can be compared with
observations of groups and clusters.

Equation~\ref{eq:field-growth-rate-specific} shows that the growth rate of
the thermal instability is a factor of 3 smaller in plasmas with heating per
unit mass than in plasmas with heating per unit volume.  More generally, the
timescale depends on the uncertain parameter $\alpha$ and cannot be directly
applied to (or inferred from) observations.  Nonetheless, it is convenient to
use $\ts{\ti} = \gr{\ti}^{-1}$ to normalize time when considering the physics
of the thermal instability with different values of $\alpha$.  We also use
the cooling time $\ts{cool} \equiv E / \scL = (\gamma \, \ocool)^{-1}$ when
we compare our results with observations.  These two timescales differ only
by an uncertain factor of order unity.

\subsection{Multi-phase Structure}
\label{subsec:multiphase}
%
\begin{figure*}
  \vfill
  \includegraphics[width=3.33in]{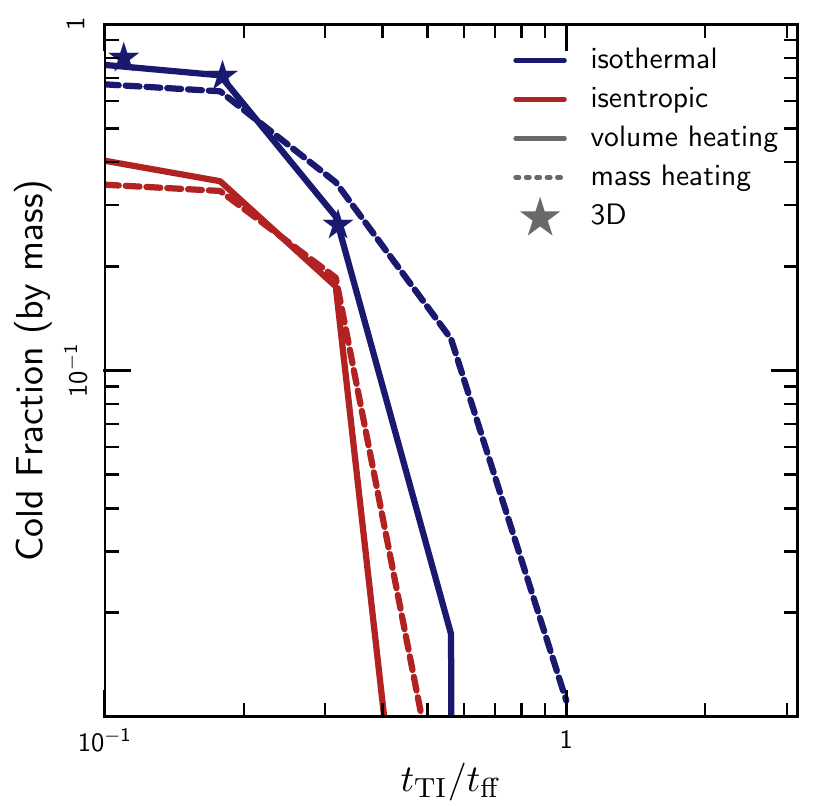}
  \hspace*{\fill}
  \includegraphics[width=3.33in]{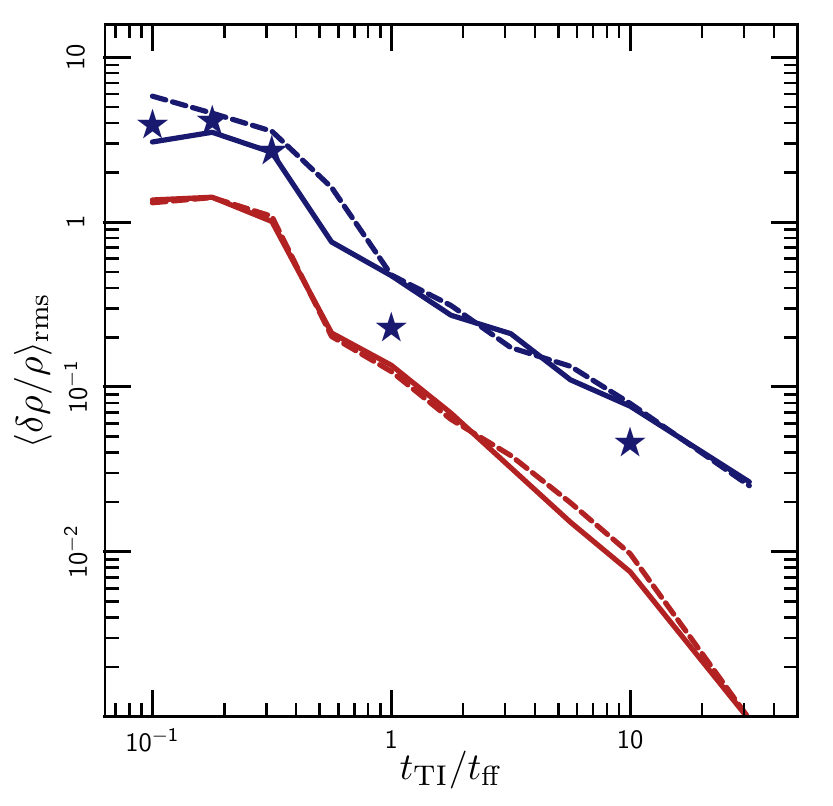}
  \caption{\textit{(Left:)} Mass fraction of cold material (with $T \leq
    T_{\mr{initial}}/3$) as a function of the timescale ratio
    $\tsri{\ti}{ff}$.  The mass in cold material drops off sharply when
    $\ts{\ti} \sim \ts{ff}$, and there is no extended multi-phase structure
    in the weak-cooling limit.  All quantities in these plots represent
    averages from $z = 0.9$--$1.1 \, H$ and from $t = 9$--$10 \, \ts{\ti}$.
    \textit{(Right:)} Fractional density inhomogeneity $\delta \rho / \rho$
    as a function of the timescale ratio $\tsri{\ti}{ff}$.  Blue (red) lines
    indicate isothermal (isentropic) initial conditions, solid (dashed) lines
    indicate volumetric (mass-weighted) heating.  The blue stars represent
    \threed simulations using our fiducial setup (isothermal initial
    condition with volumetric heating); the remaining simulations are \twod.}%
  \label{fig:saturation}
\end{figure*}
We performed simulations with the ratio of time-scales $\tsri{\ti}{ff}$
ranging from 0.1 to 10 (measured at $z = H$) and ran each for ten growth
times, until $t = 10 \, \ts{\ti}$.
\fig~\ref{fig:simulation-snapshots-detail} shows representative snapshots of
the density at the end of our fiducial simulations with volumetric heating
and isothermal initial conditions.  Our simulations show that plasmas with
cooling times shorter than the dynamical time ($\ts{\ti} \ll \ts{ff}$)
develop spatially extended multi-phase structure, whereas plasmas with
cooling times longer than the dynamical time ($\ts{\ti} \gtrsim \ts{ff}$) do
not.  Thus, the ratio $\tsri{\ti}{ff}$ controls the non-linear saturation of
the thermal instability in stratified plasmas; this observationally-testable
prediction is the primary result of our study.  This conclusion does not
depend strongly on either the initial stratification of the plasma or on our
choice of heating per unit volume.  To demonstrate this,
\fig~\ref{fig:simulation-snapshots} shows variations of our fiducial
simulations with isentropic initial conditions and with mass-weighted heating.
In all four cases, the saturated state transitions from single-phase to
multi-phase when the ratio of time-scales $\tsri{\ti}{ff}$ becomes less than
one.  Below, we describe the plasma properties in these two limits and the
physics of the transition between them.

The evolution of plasmas with short cooling times $\ts{\ti} \ll \ts{ff}$ is
straightforward: in this limit, the thermal instability develops and
saturates before the plasma can buoyantly respond.  The initial perturbations
therefore collapse into dense clumps essentially in-situ, and the \icm
develops a highly inhomogeneous, multi-phase structure wherever $\ts{\ti}(z)
< t$.  The clumps of cold gas then rain down onto the central galaxy on the
(much longer) free-fall time, while bubbles of heated gas rise outwards. The
rightmost panels of \figs~\ref{fig:simulation-snapshots-detail}
and~\ref{fig:simulation-snapshots} illustrate this behavior.  The result is a
hotter atmosphere (in which $\tsri{\ti}{ff} > 1$), filled with clumps of cold
gas.  We show in \citetalias{Sharma2011} that this end state resembles the
observed properties of some cool-core groups and clusters.

The saturation of the thermal instability is fundamentally different when the
cooling time is long compared to the dynamical time.  In this limit, gravity
and buoyancy influence the linear evolution of the perturbations (though the
growth rate changes only by a factor of two).  Nonlinearly, however, buoyancy
provides a critical saturation channel for the thermal instability that
prevents the formation of multi-phase gas.  This conclusion is qualitatively
similar to that reached by \citet{Balbus1989}; however, the physics is very
different in our case because the background atmosphere remains statistically
in thermal equilibrium for many cooling timescales.  As initial perturbations
cool and grow, they sink in the gravitational potential and mix with gas at
lower radii.  The cooling thus drives a slow, inward flow of material; the
associated mass flux is, however, significantly smaller than is predicted by
models without heating.  We return to this point in the following sections.
Rather than creating strong density inhomogeneities, cooling in this limit
excites internal gravity waves with an amplitude that depends on the
timescale ratio $\tsri{\ti}{ff}$. These waves represent the overstability
highlighted by \citet{Balbus1989} and by \citet{Binney2009}; we discuss their
saturation below.

Our results depend \textit{crucially} on the existence of a
globally-stabilizing heating mechanism; if heating were not present, the
atmospheres shown in \figs~\ref{fig:simulation-snapshots-detail}
and~\ref{fig:simulation-snapshots} would collapse monolithically.  This
globally unstable case has been studied extensively by \citet{Balbus1989}.
Consistent with their analysis, we find that atmospheres with small initial
density inhomogeneities do not form multi-phase gas, regardless of
$\tsri{\ti}{ff}$ (see \citetalias{Sharma2011} for a more detailed
discussion).

\figs~\ref{fig:simulation-snapshots-detail}
and~\ref{fig:simulation-snapshots} show that, assuming the existence of a
globally stabilizing heating mechanism, plasmas with short cooling times
$\ts{\ti} \ll \ts{ff}$ develop spatially extended multi-phase structure,
while plasmas with long cooling times do not.  The left panel of
\fig~\ref{fig:saturation} demonstrates this result more quantitatively.
Here, we plot the mass fraction of cold gas (with $T \leq
1/3\,T_{0}$) at late times in the plane $z = H$ as a
function of $\tsri{\ti}{ff}$.  (Recall that, in our units, $T_{0} \sim
T_{\mr{virial}}$.)  This figure shows that the fraction of cold gas drops
precipitously around $\tsri{\ti}{ff} \sim 1$ and that there is essentially no
multi-phase gas at large radii in simulations with $\tsri{\ti}{ff} > 1$.

The right panel of \fig~\ref{fig:saturation} quantifies the dependence of the
saturated density fluctuations on the time-scale ratio $\tsri{\ti}{ff}$ and
hints at the physics of the transition between the two limits.  Here, we plot
the root-mean-square (\rms) average of the density perturbations
\begin{align}
  \frac{\delta \rho}{\rho} \equiv \frac{\rho - \ab{\rho}}{\ab{\rho}}
\end{align}
as a function of $\tsri{\ti}{ff}$ in the plane $z = H$ as in
equation~\ref{eq:simple-heating-fn}, $\ab{\cdots}$ indicates a spatial
average at a given radius.  In the short cooling time limit, the plasma
develops multi-phase structure with large density perturbations $\delta \rho
/ \rho \gtrsim 1$.  By contrast, in the long cooling time limit, the density
perturbations saturate at much lower values $\delta \rho / \rho \ll 1$. For
plasmas with stable background stratification (\eg our isothermal initial
conditions), $\delta \rho / \rho$ in this limit represents the amplitude of
the gravity waves driven by cooling.  Note that, while the mass fraction of
cold gas drops off sharply around $\tsri{\ti}{ff} \sim 1$,
\fig~\ref{fig:saturation} shows that the mean density fluctuation is a smooth
function of this parameter, even in the weak cooling limit.  We work to
understand this quantitatively in section~\ref{sec:saturation}.

We emphasize that the difference in the evolution of plasmas with long and
short cooling times does not simply result from the thermal instability
taking longer to develop in simulations with weak cooling.  We have run each
simulation for a fixed number of growth times $\ts{\ti}$ and, if gravity were
not present, the results of our simulations with rapid and slow cooling would
be nearly identical (we have confirmed this numerically).  In fact, even in
our simulations with long cooling times, the density contrast $\delta \rho /
\rho$ becomes large near the center of the potential, where gravity is weak
(eq.~\ref{eq:gravity-defn}) and the thermal instability has time to develop.
The development of multi-phase structure depends on both gravity and cooling
and therefore on the parameter $\tsri{\ti}{ff}$, rather than simply on the
cooling time alone.

\subsection{Accreted Mass Flux}
\label{subsec:mass-flux}
%
\begin{figure}
  \includegraphics[width=3.33in]{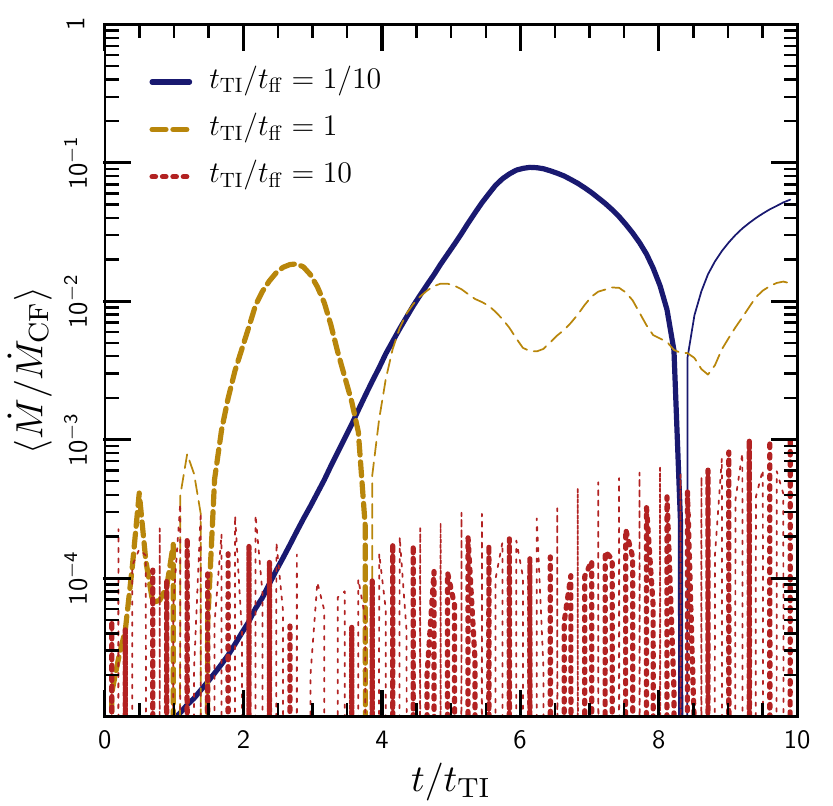}
  \caption{Mass flux (averaged from $z = 0.9$--$1.1 \, H$) as a
    function of time in \threed simulations, normalized to the cooling-flow
    flux for that atmosphere.  The mass flux is severely suppressed when
    $\ts{\ti} \gtrsim \ts{ff}$ (see eq.~\ref{eq:linear-mass-flux}).  The mass
    flux is not suppressed as strongly when $\ts{\ti} < \ts{ff}$, but it is
    highly variable.  Thick lines indicate a positive mass flux (\ie, an
    outflow), while thin lines indicate a negative mass flux (\ie, an
    inflow).  Note that gravity waves dominate the instantaneous mass flux
    when $\tsri{\ti}{ff} = 10$; the time-averaged accretion rate is much
    smaller than suggested by this plot.}%
  \label{fig:mass-flux}
\end{figure}
\fig~\ref{fig:mass-flux} shows the instantaneous, mean mass flux through the
plane $z = H$ as a function of time in three of our \threed, fiducial
simulations.  The mass fluxes are normalized to the values predicted by
cooling-flow models without heating, $\dot{M}_{\mr{CF}} = \rho H / \ts{\ti}$.
It is clear that the mass flux is strongly suppressed relative to the
cooling-flow solution whenever $\tsri{\ti}{ff} \gtrsim 1$.  This is not a
trivial consequence of our feedback heating mechanism, because for $\ts{\ti}
\lesssim \ts{ff}$, $\dot{M}$ approaches $\dot{M}_{\mr{CF}}$.  Rather, the
suppression of $\dot{M}$ for $\ts{\ti}
> \ts{ff}$ is also due to the non-linear saturation of the thermal
instability (described below).  The mass fluxes we find for $\ts{\ti} \gtrsim
\ts{ff}$ are $\lesssim 1\%$ of the cooling-flow estimates and are therefore
reasonably consistent with observational limits for cooling in the \icm
\citep{Peterson2006}.  In \citetalias{Sharma2011} we show that this
suppression is even stronger in spherical potentials and we explore its
dependence on the details of our heating model.

In the rapid cooling limit, we find that gas heated at small radii, where the
cooling time is shorter, rises up through the plane $z = H$ and initially
drives an outflow.  As the thermal instability progresses, however, this
outflow reverses and a strong accretion flow develops (although the accreted
material is all in the cold phase, rather than the hot phase; see the left
panel of \fig~\ref{fig:saturation}).  The accretion rate approaches the
cooling-flow value and eventually depletes the atmosphere of its gas.  Thus,
even our idealized feedback model (eq.~\ref{eq:heating-fn}) cannot suppress a
cooling catastrophe when $\tsri{\ti}{ff} \ll 1$.  We discuss the implications
of this result in section~\ref{sec:discussion} and, more thoroughly, in
\citetalias{Sharma2011}.

\fig~\ref{fig:mass-flux} is only meant to be suggestive, as several
subtleties in our analysis complicate a precise interpretation of the
accreted mass flux.  For instance, we use only the initial value
$\dot{M}_{\mr{CF}}$, though this quantity changes dramatically over the
course of some of our simulations.  Additionally, in simulations with
$\ts{\ti} \ll \ts{ff}$, a more appropriate normalization for the mass flux
might be $\rho H / \ts{ff}$, since the gas is not likely to flow in faster
than its free-fall rate.  \citetalias{Sharma2011} presents a much more
realistic and thorough analysis of mass accretion rates.

\subsection{Resolution Study}
\label{subsec:resolution-study}
\begin{figure*}
  \centering
  \includegraphics[width=7.0in]{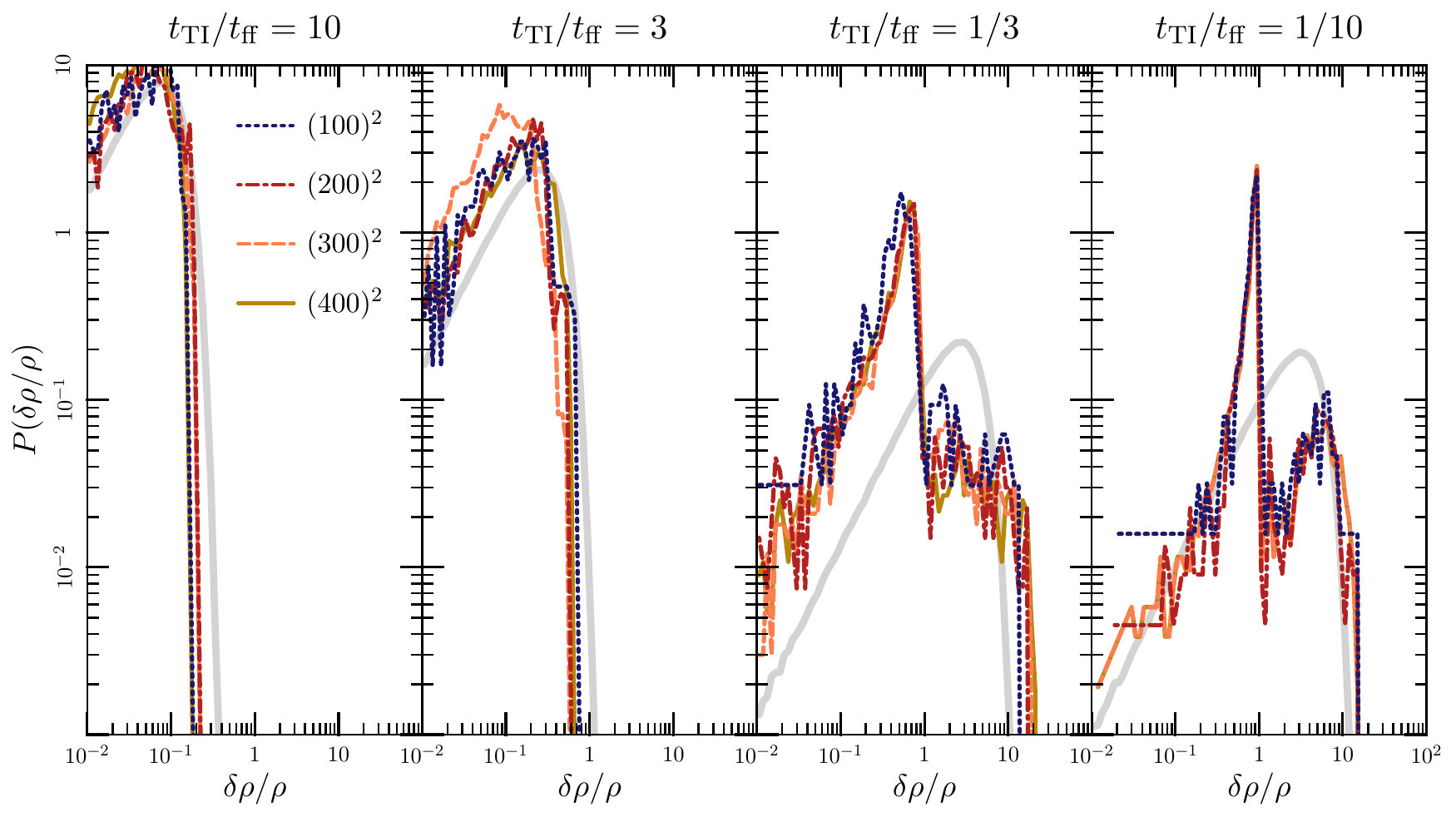}
  \caption{Convergence of the probability distribution function for
    density fluctuations $P(\delta \rho / \rho)$.  Colored lines show
    simulations at different resolutions, and the thick gray line shows the
    best-fit Gaussian distribution.  This figure shows that the density
    inhomogeneities are reasonably converged in our simulations, apart from
    the obvious fact that one can resolve finer structure, and therefore
    higher $\delta \rho / \rho$, at higher resolution.  Note that in the
    limit of rapid cooling, the properties of the high density regions are
    determined in part by the temperature floor we apply, which determines
    the density of cold clumps that can be in pressure equilibrium with the
    surrounding hot plasma.}
  \label{fig:density-pdf}
\end{figure*}
We test the numerical convergence of our results with \twod calculations on
grids of resolution $(100),\!\!\!^2$ $(200),\!\!\!^2$ $(300),\!\!\!^2$ and
$(400)^2$ for the full range of $\tsri{\ti}{ff}$.  Additionally, we performed
\threed calculations of resolution $(128)^3$ and $(256)^3$ for atmospheres
with $\tsri{\ti}{ff} = 0.1$, $1$, and $10$.  \fig~\ref{fig:density-pdf} shows
the distribution of density perturbations in our \twod simulations; this
quantity has no apparent trend with resolution.  Similarly, our \threed
simulations are nearly identical at resolutions of $(128)^3$ and $(256)^3$.
Though \fig~\ref{fig:density-pdf} only demonstrates convergence of an
integrated quantity, our simulations also ``look'' very similar at different
resolution: for example, in the rapid-cooling limit, the clumps of cold gas
have similar shapes and sizes, and they appear in the same locations.

We find rapid convergence in our simulations, even without including thermal
conduction.  By contrast, \citet{Sharma2010} found that convergence requires
resolving the Field length (\S\,\!\ref{subsec:complex-linear}) in the cold
phase of the \icm.  The temperature floor we apply (eq.~\ref{eq:cooling-fn})
implies that the Field length is not defined for the cold phase in our
simulations, and therefore that it is not relevant for convergence.  Because
the cold phase in our simulations does not cool, it can become
pressure-supported at a finite size and resist further collapse.  Convergence
is somewhat less restrictive in our simulations than in those studied by
\citet{Sharma2010}.

We performed both \twod and \threed simulations and have confirmed that they
give similar results.  Many of the plots in this paper show the results of
\twod simulations, since they are less expensive and permit a much larger
parameter study.  Because \twod simulations contain fewer grid cells than
\threed simulations, however, integrated quantities derived from \twod
calculations are noisier.  Thus, we chose to include only \threed simulations
in \figs~\ref{fig:mass-flux} and~\ref{fig:saturation-plot}.

While our \twod and \threed simulations produce similar results, they are
fundamentally different from one-dimensional simulations.  Spatial variations
between heating and cooling drive the local thermal instability in our model;
hence, the development of multi-phase structure in our simulations is an
inherently multi-dimensional effect.  Additionally, the symmetry of a
one-dimensional model prevents over-dense material from sinking and removes
an important saturation channel from the thermal instability
(\S\,\!\ref{sec:saturation}).  Much of the physics we describe in this paper
is therefore absent in one-dimensional treatments of the \icm such as those
described in \citet{Ciotti2001} and \citet{Guo2008}.

\subsection{Sensitivity to the Heating Function}
\label{subsec:heating-sensitivity}
%
\begin{figure}
  \centering
  \includegraphics[width=3.33in]{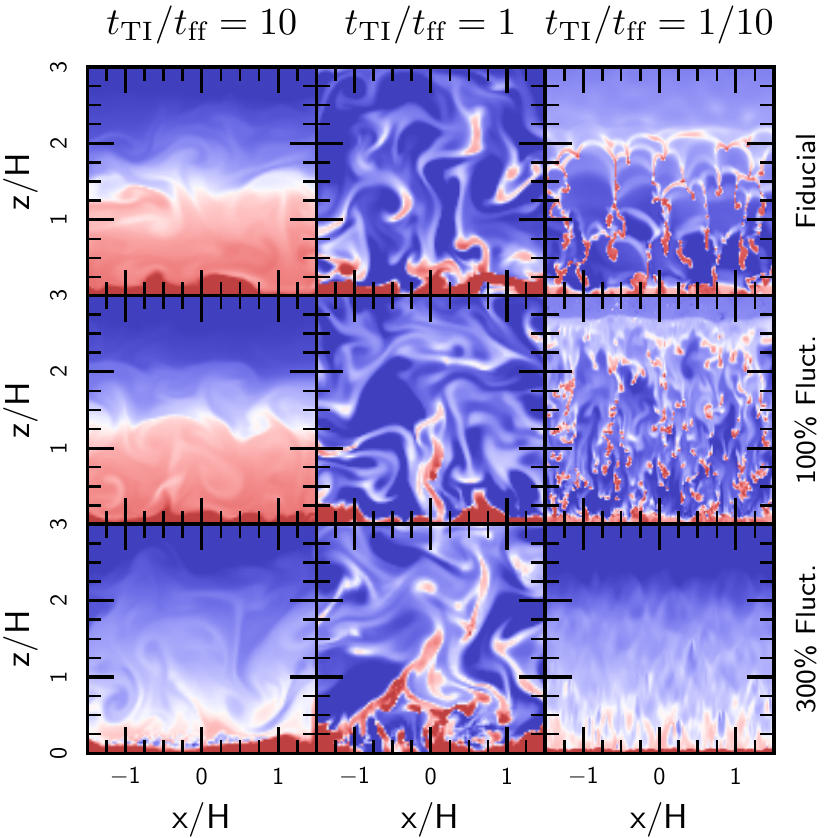}
  \caption{Comparison of the plasma density in simulations with our fiducial
    heating function (equation~\ref{eq:heating-fn}) to simulations where we
    have added significant, random fluctuations to the heating function \scH
    (see eq.~\ref{eq:fluct-heating}).  These are white noise fluctuations
    with a temporal correlation $\ts{corr} = \ts{\ti}$; simulations with
    longer correlation times $\ts{corr} = 10\, \ts{\ti}$ give similar
    results (see \fig~\ref{fig:mass-T-hist}).  From top to bottom, the
    panels show the density at $t = 10 \, \ts{\ti}$ in our fiducial
    simulations, simulations with $100\%$ fluctuations in heating, and
    simulations with $300\%$ fluctuations in heating.  Fluctuations of
    $300\%$ produce a cooling flow, but $100\%$ fluctuations do not and
    instead produce results similar to our fiducial model.  In all panels,
    color represents the log of the density, which ranges from $10^{-2}$
    (blue) to $10$ (red). }%
  \label{fig:heating-fluctuations}
\end{figure}
%
\begin{figure*}
  \centering
  \includegraphics[width=7.0in]{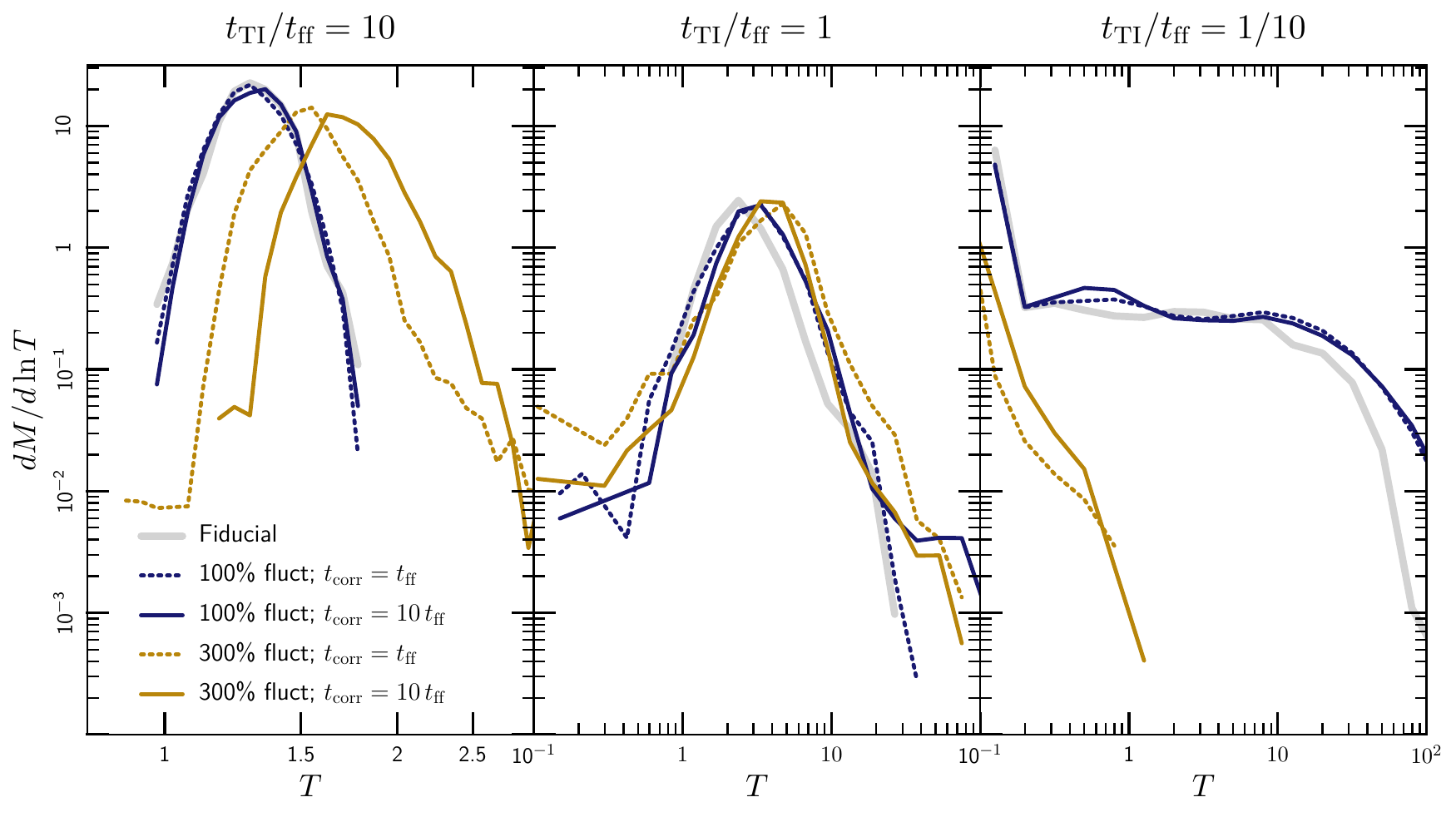}
  \caption{Gas mass as a function of temperature in simulations with
    different types of fluctuations about thermal equilibrium, measured at
    $t = 10 \ts{\ti}$ and $z \sim H$.  These simulations are for isothermal
    initial conditions in which the initial temperature $T = 1$.  The
    fluctuations are of the form $\scH \rightarrow \scH (1+\delta)$, where
    $\delta(\vec{x},t)$ has a white-noise spatial spectrum and a temporal
    coherence time $\ts{corr}$ (see \S\,\!\ref{subsec:heating-sensitivity}
    for details).  The gas properties are not sensitive to strong
    fluctuations in heating of up to 100\% in amplitude.  Stronger
    fluctuations of 300\% generate significant cold material when
    $\tsri{\ti}{ff} \gg 1$; the cold material sinks to small radii (see
    \fig~\ref{fig:heating-fluctuations}), leading to a modest heating of the
    gas that remains at $z \sim H$.  When $\tsri{\ti}{ff} \ll 1$,
    fluctuations of 300\% break our ansatz of approximate thermal
    equilibrium and induce a cooling catastrophe.  The correlation time
    $\ts{corr}$ has only a modest influence on these results. }%
  \label{fig:mass-T-hist}
\end{figure*}
An important test of our model is the sensitivity of our conclusions to the
details of the (unknown) heating function.  We study this dependence by
adding random heating fluctuations of the form
\begin{align}
  \scH \rightarrow \scH \left(1+\delta\right),\label{eq:fluct-heating}
\end{align}
where $\delta(\vec{x},t)$ is a Gaussian-random field with a white-noise
spatial power spectrum and a temporal autocorrelation function
$R_{\delta\delta}(\tau) = e^{-\tau/\ts{corr}}$.  Thus, $\delta$ introduces
both spatial and temporal imbalances between heating and cooling, which
persist for the coherence time $\sim\ts{corr}$.  These fluctuations detune
our feedback model while still preserving average thermal equilibrium, and
are intended to mimic the temporal and spatial differences between heating
and cooling which might arise in a more realistic feedback scenario. More
importantly, including these fluctuations allows us to distinguish between
results that are a consequence of the exact (and, in detail, unphysical)
balance in equation~\ref{eq:heating-fn} and results that are more robust and
are primarily a consequence of global thermal stability.

We carried out \twod simulations with $\ts{corr} = \{0.1$, $1$, $10\} \times
\ts{ff}$ and with the fluctuations normalized to root-mean-square (\rms)
amplitudes of 50\%, 100\% and 300\%.  (Note that we quote the \rms, or
`1$\sigma$' amplitude of the fluctuations; the peak values are considerably
higher.)  \fig~\ref{fig:heating-fluctuations} shows images of the density
fluctuations for these simulations and \fig~\ref{fig:mass-T-hist} shows the
temperature distribution function for different values of $\ts{corr}$ and the
fluctuation amplitude.  These figures demonstrate that our conclusions about
the development of the thermal instability are essentially unaffected by
order-unity fluctuations, over at least 10 cooling times.  This important
result implies that, as long as the plasma is in approximate global thermal
equilibrium on reasonable time-scales $\sim \ts{\ti}$ and length-scales $\sim
H$, the development and saturation of local thermal instability will proceed
approximately as shown in
\figs~\ref{fig:simulation-snapshots-detail}--\ref{fig:mass-flux}.  We think
that the existence of an approximate thermal equilibrium, rather than the
specific details of our heating function (eq.~\ref{eq:heating-fn}),
determines how the thermal instability develops and saturates.  This
conclusion is bolstered by \citetalias{Sharma2011}, which finds very similar
results using an entirely different heating function.

Figs~\ref{fig:heating-fluctuations} and~\ref{fig:mass-T-hist} show that
extremely strong heating fluctuations with \rms amplitudes of 300\% spoil the
thermal equilibrium of the plasma and induce a cooling catastrophe; even our
extremely optimistic feedback model cannot withstand arbitrarily large
heating perturbations.  Though the feedback mechanism is not yet understood
in clusters, this places a constraint on the heating: it should not differ
persistently from the local cooling rate by more than a factor of
several. \fig~\ref{fig:mass-T-hist} shows that this conclusion is essentially
independent of the coherence time $\ts{corr}$ of the heating.

\section{Interpretation of the Non-Linear Saturation}
\label{sec:saturation}
%
\begin{figure}
  \includegraphics[width=3.33in]{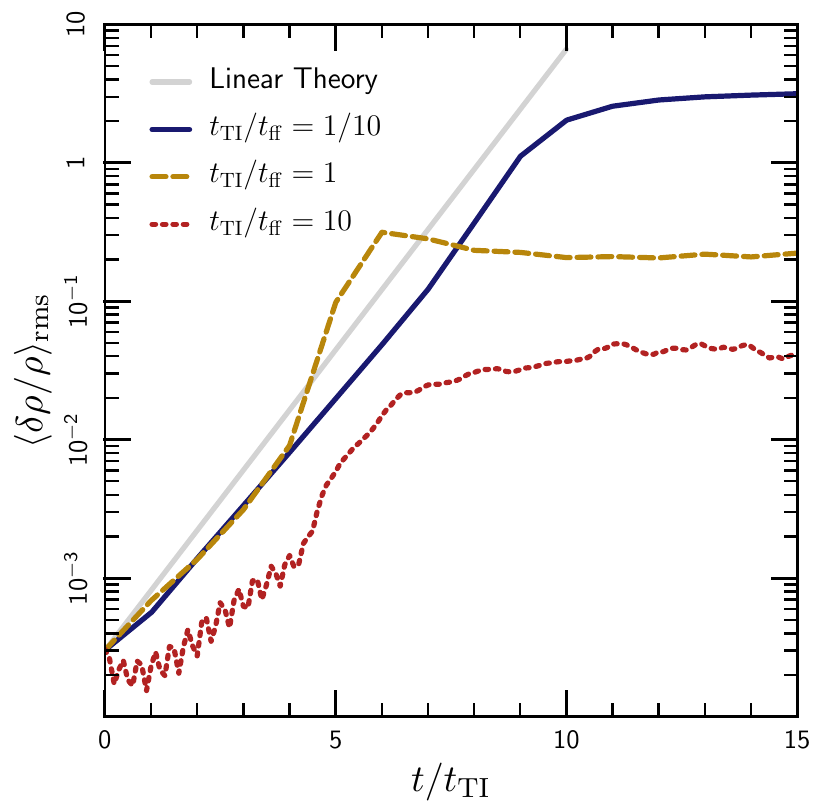}
  \caption{Evolution of the density fluctuation
    $\delta \rho / \rho$ as a function of time in our simulations with
    isothermal initial conditions.  The plotted quantity is an \rms average
    from $z = 0.9 H$--$1.1 H$.  The density inhomogeneity grows from the
    initial perturbation until the characteristic infall time becomes
    comparable to the local cooling time.  At this point, the density
    contrast saturates at approximately the value given by
    equation~\ref{eq:analytic-expression-iso}.  This figure shows the results
    from \threed simulations, but the results from \twod simulations are
    similar. }%
  \label{fig:saturation-plot}
\end{figure}
In this section, we show that the linearized dynamical equations provide
valuable insight into the non-linear saturation of the thermal instability
and its astrophysical implications.  As in
section~\ref{sec:simulation-results}, we focus on the development of
multi-phase structure (\S\,\!\ref{subsec:multiphase-interpretation}) and on
the accreted mass flux (\S\,\!\ref{subsec:massflux-interpretation}), which
have been extensively studied observationally.  Our basic procedure is to
estimate a saturation amplitude for the linear instability.  Because we use
linearized equations, the interpretation in this section only strictly holds
in the weak cooling limit ($\ts{\ti} \gg \ts{ff}$), so that the density
perturbations remain relatively small.

\fig~\ref{fig:saturation-plot} illustrates the development and saturation of
the thermal instability.  The density inhomogeneity $\delta \rho / \rho$ (or
any other quantity linear in the perturbation) initially grows exponentially
according to the dispersion relation (eq.~\ref{eq:linear-dispersion}), but
eventually freezes out at a finite amplitude.  This amplitude, along with the
relations~\ref{eq:pert-mass}--\ref{eq:pert-e}, then approximately determines
the state of the plasma at late times.  In the following sections, we
estimate this amplitude and show that we can reproduce elements of the
non-linear saturation shown in \fig~\ref{fig:saturation}.
\fig~\ref{fig:saturation-plot} shows that the difference between atmospheres
which develop multi-phase gas and ones which do not is \emph{fundamentally a
  non-linear effect}.  The linear growth rate of the perturbations is largely
independent of the time-scale ratio \tsri{\ti}{ff}, and it is the saturation
which determines the degree of inhomogeneity at late times.

\subsection{Saturation Amplitudes}
\label{subsec:saturation-interpretation}
In the limit that the plasma is buoyantly neutral ($N = 0$), we can estimate
the saturation amplitude by inspecting the linearized, Lagrangian form of the
momentum equation (eq.~\ref{eq:consmom}):
\begin{align}
  \frac{d v_z}{d t} = -\frac{\delta n}{n} g .\label{eq:perturbed-mom}
\end{align}
The characteristic inflow (or outflow) time for a perturbed fluid element is
$\ts{sink} \sim H/\delta v_z$.  Initially, $\delta v_z$ is small and this
inflow time is long compared to the growth time of the thermal instability.
As the perturbation grows, however, $\delta n / n$ increases and the fluid
element accelerates according to equation~\ref{eq:perturbed-mom}. The inflow
time $\ts{sink}$ thus becomes shorter as the instability develops.  We assume
that the growth ceases when the inflow time is comparable to, or slightly
shorter than, the growth time of the thermal instability.  In this case, the
thermal instability saturates when the velocity satisfies
\begin{align}
  \delta v_z &\sim \frac{H}{\ts{\ti}} .\label{eq:saturated-velocity}
\end{align}
Thus, non-linear saturation occurs when a fluid element flows to smaller
radii after one cooling time, as seems intuitively reasonable.

The physical picture of a sinking fluid element does not apply in a stably
stratified atmosphere, since the fluid element does not flow monotonically
inwards, but instead oscillates with the gravity wave frequency $N$.  The
velocity associated with this oscillation dwarfs the mean, inward velocity.
These waves are sourced by cooling, however, and we assume that they reach a
steady-state in which the dissipation rate due to non-linear mode coupling
equals the driving rate due to the thermal instability $\sim\ts{\ti}^{-1}$.
Thus the instability saturates when the dissipation time $\ts{diss} \sim H /
\delta v \sim \ts{\ti}$, where we have assumed strong turbulence and used the
fact that the waves are driven on large scales, $\sim~H$ (as suggested by the
bottom panels in \fig~\ref{fig:simulation-snapshots-detail}).  Though this
saturation mechanism is very different from that described above for
buoyantly neutral plasmas, it implies an equivalent saturation amplitude.  We
therefore assume that equation~\ref{eq:saturated-velocity} describes the
late-time evolution of the perturbations in all of our simulations.  We show
in the following sections how the behavior described in
section~\ref{sec:simulation-results} can be understood in terms of this
saturation amplitude.

\subsection{Multi-Phase Structure}
\label{subsec:multiphase-interpretation}
Inserting our ansatz for the saturation amplitudes
(equation~\ref{eq:saturated-velocity}) into the momentum
equation~\eqref{eq:pert-mom} and using the dispersion relation
(equation~\ref{eq:linear-dispersion}) to replace $\omega$, we express the
density inhomogeneity $\delta n / n$ at late times in terms of other
properties of the plasma:
\begin{align}
  \frac{\delta n}{n} \propto
  \left(\tsr{ff}{\ti}\right)^2
  \left[ 1 \mp \sqrt{1 - 4 \kperp^2 \left(\tsr{\ti}{buoy}\right)^2} \right]
  \label{eq:lin-inhomo}
\end{align}
where $\ts{buoy} \equiv N^{-1}$ and $\kperp^2 \equiv (1 - \oldhat{k}_z^2)$ is
the squared horizontal component of the direction of the wave vector
(typically $\sim 1$).

Equation~\ref{eq:lin-inhomo} has the asymptotic forms
\begin{subequations}
  \begin{align}
    \frac{\delta n}{n} &\propto \left(\tsr{ff}{\ti}\right)^2
    & \ts{\ti} \ll \ts{buoy} \label{eq:analytic-expression-ise} \\
    \frac{\delta n}{n} &\propto
    \left(\tsr{ff}{\ti}  \right)
    \left(\tsr{ff}{buoy}\right)
    & \ts{\ti} \gg \ts{buoy} .\label{eq:analytic-expression-iso}
  \end{align} \label{eq:analytic-expectation}
\end{subequations}
Equation~\ref{eq:analytic-expectation} shows that weakly stratified plasmas
with $\ts{\ti} \ll \ts{buoy}$ develop smaller density inhomogeneities than
plasmas with $\ts{\ti} \gg \ts{buoy}$ (in the limit that $\ts{\ti} \gtrsim
\ts{ff}$).  This difference arises because because plasmas with $\ts{\ti} \gg
\ts{buoy}$ can sustain internal gravity waves, while atmospheres with
$\ts{\ti} \ll \ts{buoy}$ cannot.

Somewhat surprisingly, the right panel of \fig~\ref{fig:saturation} shows
that the measured dependence of $\delta n / n$ on the time-scale ratio
$\tsri{\ti}{ff}$ is in good agreement with
equation~\ref{eq:analytic-expression-iso} for both isothermal and isentropic
initial conditions, even though $\ts{buoy} \rightarrow \infty$ in an
isentropic atmosphere.  This is because
equation~\ref{eq:analytic-expression-ise} applies only when the atmosphere is
very nearly buoyantly neutral (at least when $\ts{\ti}$ is long compared to
the dynamical time, as must be for equation~\ref{eq:analytic-expectation} to
be valid).  While this is the case initially in our isentropic atmospheres,
the entropy gradient evolves somewhat with time and
equation~\ref{eq:analytic-expression-ise} ceases to describe the plasma after
only a few cooling times.  The saturated values of $\ts{buoy}$ differ in our
simulations with isentropic and isothermal initial conditions;
equation~\ref{eq:analytic-expression-iso} suggests that this may explain the
systematic offset between these simulations shown in
\fig~\ref{fig:saturation}.  For the longest cooling times, the evolution of
the background profile is smallest; the slight steepening of $\delta \rho /
\rho$ for our isentropic simulations in this limit may represent an
intermediate case between equations~\ref{eq:analytic-expression-ise}
and~\ref{eq:analytic-expression-iso}.

\subsection{Accreted Mass Flux}
\label{subsec:massflux-interpretation}
We can also use our estimate of the non-linear saturation to understand the
inward mass flux induced by the thermal instability.  Defining the mass flux
$\dot{M} = \ab{\delta n \, \delta v_z}$ and using the estimates of $\delta v$
and $\delta n/n$ from the momentum equation and from
equation~\ref{eq:lin-inhomo}, we find
\begin{align}
  \begin{split}
    \dot{M}
    = &\dot{M}_{\mr{CF}}
    \left( \tsr{ff}{\ti} \right)^2 \frac{1}{4 \kperp^2} \\
    &\times \left<
      \mr{Re}\left(\left[ 1 +
          \sqrt{1 - 4 \kperp^2 \left(\tsr{\ti}{buoy}\right)^2 }
        \right]e^{i \vec{k}\cdot\vec{x}}\right)
      \mr{Re}\left(e^{i \vec{k}\cdot\vec{x}}\right) \right>,
  \end{split}
\end{align}
where $\dot{M}_{\mr{CF}} = \rho H / \ts{\ti}$ is the mass flux expected in
the absence of heating (recall that we are in the limit that $\ts{\ti} \gg
\ts{buoy}$). This yields
\begin{align}
  \frac{\dot{M}}{\dot{M}_{\mr{CF}}} &\simeq
  \begin{dcases}
    \frac{1}{4} \left(\tsr{ff}{\ti}\right)^2
    - \frac{1}{4} \left(\tsr{ff}{buoy}\right)^2
    & \ts{\ti} \ll \ts{buoy} \\
    \frac{1}{8} \left(\tsr{ff}{\ti}\right)^2
    & \ts{\ti} \gg \ts{buoy}
  \end{dcases} . \label{eq:linear-mass-flux}
\end{align}
Equation~\ref{eq:linear-mass-flux} shows that the mass flux is dramatically
suppressed when $\tsri{\ti}{ff} \gg 1$, qualitatively consistent with
\fig~\ref{fig:mass-flux}.  Furthermore, this suppression is nearly
independent of the initial stratification of the plasma, even though stably
stratified plasmas show much stronger density inhomogeneities
(eq.~\ref{eq:analytic-expectation}).  This is because the internal gravity
waves that enhance $\delta n / n$ when $\tsri{\ti}{buoy} \gg 1$ do not
contribute to the net mass flux.  Note that gravity waves dominate the
instantaneous mass flux shown in \fig~\ref{fig:mass-flux} when
$\tsri{\ti}{ff} \gg 1$; the time-averaged accretion rate is smaller than the
figure suggests.

\section{Simulations Including Conduction and Magnetic Fields}
\label{sec:complex-sims}
%
\begin{figure*}
  \centering
  \includegraphics[width=7.0in]{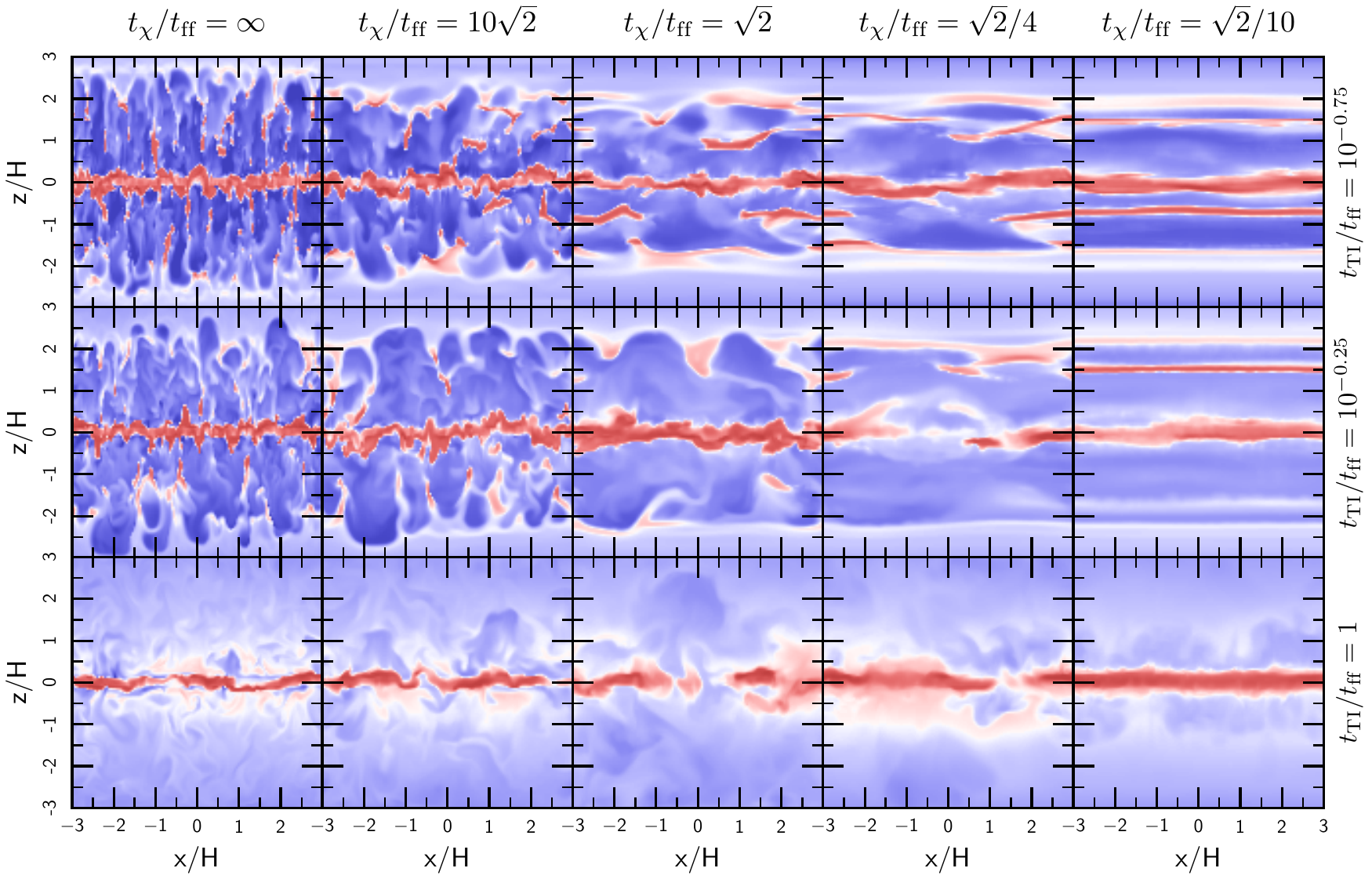}
  \caption{Comparison of the gas density in the non-linear state of
    simulations with different time-scale ratios $\tsri{\ti}{ff}$ and
    $t_{\chi} / \ts{ff}$, including an initially horizontal magnetic field
    and anisotropic thermal conduction.  We take the conduction time to be
    the time it takes heat to diffuse across one scale-height: $t_{\chi}
    \equiv H^2 / \chi$.  Rapid conduction dramatically changes the
    morphology of the cold gas, smearing it out in the direction of the
    magnetic field.  However, conduction does not appreciably change the
    mass of gas in the cold phase (\fig~\ref{fig:cold-fraction-conduction}).
    The ratio of timescales $\tsri{\ti}{ff}$ still determines whether or not
    the plasma develops multi-phase structure, even in the limit of rapid
    conduction.  In all frames, the color scale represents the log of the
    density; blue corresponds to a density of $10^{-2}$ and red corresponds
    to a density of $10$.  The filaments are very straight in simulations
    where the Field length is longer than the domain size (right-most
    column) because conduction effectively eliminates all horizontal
    structure in the initial perturbations; the subsequent evolution is
    therefore nearly two-dimensional.  We show both the top and bottom of
    the computational domain here to emphasize that the number of
    filaments/blobs produced by the thermal instability is somewhat
    stochastic.}%
  \label{fig:conduction-morphology}
\end{figure*}
%
\begin{figure}
  \centering
  \includegraphics[width=3.33in]{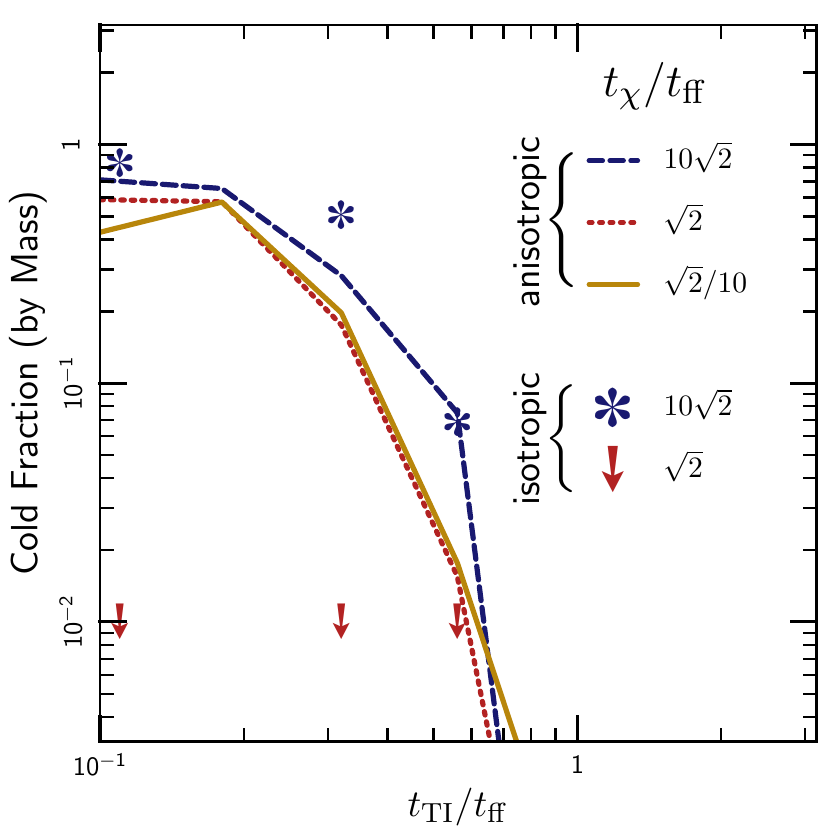}
  \caption{Mass fraction of cold material (with $T \leq T_0/3$) as a
    function of the timescale ratio $\tsri{\ti}{ff}$, for simulations with
    different conductivities.  As in \fig~\ref{fig:saturation}, this mass
    fraction is determined by averaging from $z = 0.9$--$1.1 \, H$ and from
    $t = 9$--$10 \, \ts{\ti}$.  Lines show simulations with anisotropic
    thermal conduction and points indicate simulations with isotropic
    conduction.  As suggested by \fig\,\!\ref{fig:conduction-morphology},
    anisotropic thermal conduction does not strongly influence the
    multi-phase structure produced by the thermal instability: though the
    conductivities in these simulations differ by a factor of 100, the cold
    mass fractions agree to within about a factor of 2.  In particular, the
    results appear to converge in the rapid-conduction limit.  By contrast,
    we see no multi-phase structure in simulations with isotropic conduction
    if the Field length is comparable to, or greater than a scale-height
    (the arrows in this figure indicate upper limits on the mass fractions
    of cold gas). }%
  \label{fig:cold-fraction-conduction}
\end{figure}
The previous sections describe a simplified model of the thermal instability
that neglects both conduction and the dynamical effect of the magnetic field.
This model nicely isolates the physics of the thermal instability, but
astrophysically it is too idealized.  For example, conduction is critical for the
thermal evolution of the plasma over a wide range of scales in the \icm, and
\citet{Balbus2000} and \citet{Quataert2008} have shown that this completely
changes the stability and dynamics of the plasma.  In this section, we
present results including magnetic fields and conduction and show that the
conclusions from the previous sections largely apply in this more realistic
case.

\subsection{Setup}
\label{subsec:complex-setup}
Our setup is very similar to that described in sections~\ref{sec:method}
and~\ref{sec:setup}.  We generalize equations~\ref{eq:consmom}
and~\ref{eq:inte} to include the effects of thermal conduction and the
magnetic field:
\begin{align}
  \frac{\partial}{\partial t} \left( \rho \, \vec{v} \right)
  + \nabla \cdot \Bigg[ \rho \, \vec{v}\otimes\vec{v} +
  \bigg(P + \frac{B^2}{8\pi}\bigg) \mathbfss{I} +
  \frac{\vec{B} \otimes \vec{B}}{4\pi} \Bigg]
  = \rho \, \vec{g} ,\tag{\ref{eq:consmom}$'$}\label{eq:consmom-cond} \\
  \rho \, T \frac{d s}{d t} = \left(\scH - \scL\right) -
  \nabla \cdot \Qcond ,\tag{\ref{eq:inte}$'$}\label{eq:inte-cond}
\end{align}
where $\vec{B}$ is the magnetic field and $\Qcond$ is the conductive heat
flux.  We evolve the magnetic field using the induction equation:
\begin{align}
  \vphantom{\frac{\Sigma}{\Sigma}}
  \frac{\partial \vec{B}}{\partial t}&
  = \nabla \times (\vec{v} \times \vec{B})\tag{1d$'$}\label{eq:consflux}.
\end{align}\label{eq:cond-dyn}
We have ignored both (explicit) viscous and magnetic dissipation in
equations~\ref{eq:consmom-cond}--\ref{eq:consflux}.  These effects can
influence \mhd simulations in subtle and unexpected ways \citep[see,
\eg][]{Fromang2007,Davis2010}, and so will need to be studied in detail in
the future.

The thermal conductivity of the plasma is strongly anisotropic in the \icm
and as a result the conductive heat flux is given by \citep{Braginskii1965}
\begin{align}
  \Qcond = -n \kb \chie \, \hat{b} \,
  (\hat{b} \cdot \nabla T),\label{eq:cond-flux}
\end{align}
where $\hat{b} = \vec{B} / B$ is a unit vector in the direction of the
magnetic field and $\chie$ is the thermal diffusivity of free electrons (with
units of cm$^{2}$/s).  While the diffusivity $\chie$ depends sensitively on
temperature \citep{Spitzer1962}, we take it to be constant in this
exploratory analysis.  This enables us to control the ratio of the conduction
time to other timescales in the problem and thus to isolate the physics of
cooling and conduction.  Note that we still use the heating function defined
by equation~\ref{eq:heating-fn}; any conductive heating or cooling of the
plasma happens on top of the feedback heating.

We initialize the plasma with a weak, horizontal magnetic field.  (By `weak,'
we mean that magnetic tension is negligible in our simulations.)
Because we impose reflecting boundary conditions at the upper and lower
boundaries of the domain (\S\,\!\ref{sec:setup}), the magnetic field remains
horizontal there and prohibits a conductive heat flux into the domain.

As before, we solve equations~\ref{eq:consmass}
and~\ref{eq:consmom-cond}--\ref{eq:consflux} using \textsc{Athena} with the
modifications described in section~\ref{sec:setup}.  We also implement
equation~\ref{eq:cond-flux} via operator splitting, using the anisotropic
conduction algorithm described in \citet{Parrish2005} and \citet{Sharma2007}.
In particular, we use the monotonized central difference limiter on
transverse heat fluxes to ensure stability.  This conduction algorithm is
sub-cycled with respect to the main integrator with a time step $\Delta t
\propto (\Delta x)^{2}$; these simulations are therefore more computationally
expensive than adiabatic \mhd calculations, especially at high resolution.

\subsection{Linear Properties}
\label{subsec:complex-linear}
We linearize equations~\ref{eq:consmass}
and~\ref{eq:consmom-cond}--\ref{eq:consflux} and perform a WKB analysis
(see~\citealt{Quataert2008} for more details).  Assuming that magnetic
tension is negligible, and proceeding as in section~\ref{subsec:stability},
the dispersion relation for the plasma is \citep[\cf][]{Balbus2010}
\begin{align}
  p^3 - p^2 \gr{\textsc{f}} + p \, N^2 \kperp^2 - \ocond \, \gr{\hbi}^2 = 0 .
\end{align}
In the above, $p = -i \omega$ is the growth rate of the perturbation,
$\gr{\textsc{f}} = \gr{\ti} - \ocond$ is the growth rate of the thermal
instability accounting for conduction \citep{Field1965},
\begin{align}
  \ocond = \frac{\gamma-1}{\gamma} \chie
  \left(\hat{b} \cdot \vec{k}\right)^2
\end{align}
is inversely proportional to the conduction time across the wavelength of the
perturbation, and
\begin{align}
  p^2_{\mr{\hbi}} = g \partiald{\,\ln T}{z} \, \times
  &\left[ (2 \oldhat{b}_z^2 - 1)(1 - \oldhat{k}_z^2) -
    2 \oldhat{b}_x \oldhat{b}_z \oldhat{k}_x \oldhat{k}_z
  \right] \label{eq:hbi-growth-rate}
\end{align}
is the growth rate of either the magnetothermal instability
\citep[\mti;][]{Balbus2000}, or the heat-flux driven buoyancy instability
\citep[\hbi;][]{Quataert2008}.

The \mti is unlikely to influence the development of multi-phase structure in
galaxy clusters, since it operates outside the cool core, where the ratio of
timescales $\tsri{\ti}{ff}$ is typically much greater than unity.  While the
\hbi does operate efficiently in cool cores, it behaves like ordinary stable
stratification in its saturated state \citep{McCourt2011} and the growth time
$\ts{\hbi}$ is analogous to the timescale $\ts{buoy}$ used earlier.  Thus, we
do not expect the \hbi or \mti to change our results in any essential way
(although this must be studied more carefully in future work).  We anticipate
that the same will be true for the overstabilities associated with the \mti
and \hbi \citep{Balbus2010}.  In this section, we use simulations with
isothermal initial conditions (in which $p_{\mr{\hbi}} \rightarrow 0$) so
that these instabilities and overstabilities do not operate (at least in our
initial conditions).  This allows us to focus on the physics of thermal
instability.

The conduction frequency $\ocond$ is a function of scale, while the growth
rate of the thermal instability $\gr{\ti}$ is not; the modified growth rate
$\gr{\textsc{f}}$ therefore must switch sign at the length-scale
\begin{align}
  \lambda_{\mr{F}} = |\hat{b}\cdot\hat{k}| \times \left[
    (2\pi)^2 \frac{\gamma-1}{\gamma} \frac{\chi_{\mr{e}}}{\gr{\ti}}
  \right]^{1/2},\label{eq:field-length}
\end{align}
known as the Field length \citep{Field1965}.  Intuitively, the Field length
is the distance heat can diffuse in one cooling time; if the wavelength of a
perturbation is larger than this distance, conduction cannot stabilize it
against cooling and the perturbation grows exponentially.

Conduction suppresses the thermal instability on scales smaller than the
Field length, but the Field length in a magnetized medium depends on
direction, as well as position.  Even if the term in square brackets in
equation~\ref{eq:field-length} becomes arbitrarily large, the Field length
will be small in directions orthogonal to the magnetic field.  Because of
this anisotropy, the thermal instability can still grow on scales much
smaller than $\sqrt{\chi_{\mr{e}} \ts{\ti}}$.  \citet{Sharma2010} have
studied this growth in the absence of gravity; here we generalize their
results to stratified plasmas.

The growth rate for the thermal instability in our simulations is
\begin{subequations}
  \begin{align}
    p &= \frac{1}{2}
    \left[ \gr{\textsc{f}} \pm \sqrt{\gr{\textsc{f}}^2 - 4 N^2 \kperp^2} \right] \\
    &= \begin{dcases}
      \gr{\textsc{f}} - N^2 \kperp^2 / \gr{\textsc{f}} & \gr{\textsc{f}} \gg |N| \\
      \frac{1}{2}\gr{\textsc{f}} \pm i N \kperp        & \gr{\textsc{f}} \ll |N|
    \end{dcases} .
  \end{align}\label{eq:cond-growth-rate}
\end{subequations}
Equation~\ref{eq:cond-growth-rate} shows that the characteristic growth time
of the thermal instability is $\gr{\textsc{f}}^{-1}$, regardless of the
entropy gradient.  The growth rate $\gr{\textsc{f}}$ reduces to $\gr{\ti}$ on
large scales; thus, our results with and without conduction are very similar
on scales larger than the Field length.  Conduction prevents perturbations
from growing below the Field length and therefore plays a similar role to the
temperature floor in our non-conducting simulations.  The primary difference
between our conducting and non-conducting simulations is that the Field
length is anisotropic in the conducting simulations, and the thermally
unstable fluid elements collapse into long filaments, rather than the
approximately spherical clumps shown in
\figs~\ref{fig:simulation-snapshots-detail}
and~\ref{fig:simulation-snapshots}.

\subsection{Numerical Results}
\label{subsec:complex-results}
\fig~\ref{fig:conduction-morphology} shows \twod slices of \threed
simulations with different values of the cooling constant $\Lambda_0$ and the
conductivity $\chie$.  We use only \threed simulations in this section
because, just as an over-dense fluid element cannot sink in one dimension,
the dynamics of a sinking magnetized filament changes in going from two to
three dimensions.  These simulations all use our isothermal initial condition
and initially have weak, horizontal magnetic field lines in the plane of the
figure.  Rapid conduction smears out the cold clumps into filaments of length
$\sim \lambda_{\mr{F}}$, but does not otherwise alter the growth of the
thermal instability.  Specifically, \fig~\ref{fig:conduction-morphology}
demonstrates that, even in the limit of very rapid conduction, the ratio of
time-scales $\tsri{\ti}{ff}$ determines whether the plasma develops
multi-phase structure.  This result depends \textit{critically} on the
anisotropic nature of thermal conduction.  In the rightmost panels of
\fig~\ref{fig:conduction-morphology}, the Field length is larger than the
entire simulation domain; if conduction were isotropic, the entire atmosphere
would become nearly isothermal and the thermal instability would be
suppressed.  The insulating effect of the magnetic field permits large
temperature gradients orthogonal to the magnetic field and thus the formation
of multi-phase structure \citep{Sharma2010}.

\fig~\ref{fig:cold-fraction-conduction} quantifies the effect of conduction
on the thermal instability: we show the mass fraction of cold gas (as in the
left panel of \fig~\ref{fig:saturation}) for \threed simulations with
different thermal conductivities.  In simulations with anisotropic thermal
conduction, this mass fraction is almost independent of the conductivity, and
it appears to converge in the limit that the conductivity becomes large.
This behavior is consistent with \fig~\ref{fig:conduction-morphology}.
Together, these results imply that anisotropic conduction alters the
morphology of the gas in the cold phase, but not the presence, absence, or
amount of multi-phase structure.

We have also run a number of simulations with isotropic thermal conduction.
These simulations use the same setup as before, but with the conductive heat
flux $\Qcond = - n \kb \chie \nabla T$, where (as before) $\chie$ is a
constant, free parameter.  In order to prevent conduction from changing the
total energy content of the plasma, we set $\chie = 0$ at the upper and lower
boundaries of the computational domain so that there is no conductive heat
flux into the domain.  \fig~\ref{fig:cold-fraction-conduction} shows that,
while anisotropic conduction does not strongly influence the amount of cold
gas produced by the thermal instability, isotropic conduction can quench it
entirely: we see no multi-phase structure in our simulations with isotropic
conduction whenever the Field length is comparable to, or larger than, the
pressure scale-height.  These conclusions also apply to other properties of
the plasma quantified in section~\ref{sec:simulation-results}, \eg the
accreted mass flux: anisotropic thermal conduction has little effect on this
quantity, while isotropic thermal conduction can strongly suppress~it.

\citet{Voit2008} suggested that thermal instability produces multi-phase gas
in clusters when the Field length is comparable to, or smaller than, the size
of the cool core, but that conduction suppresses the formation of multi-phase
structure for larger Field lengths.  Coincidentally, in typical cool-core
clusters, this criterion is quantitatively similar to our criterion on the
ratio $\tsri{\ti}{ff}$.\footnote{This comparison makes use of the result from
  \citetalias{Sharma2011} that the threshold for multi-phase gas in spherical
  systems is closer to $\tsri{\ti}{ff} \sim 10$.}  However, because the \icm is
magnetized, thermal conduction is extremely anisotropic; the results of this
section demonstrate that even very rapid thermal conduction cannot suppress
local thermal instability.  Thermal conduction only stabilizes modes parallel
to the magnetic field, and multi-phase structure continues to develop via
perturbations that are roughly orthogonal to the local magnetic field.

\section{Discussion}
\label{sec:discussion}
\begin{figure*}
  \centering
  \includegraphics[width=7.0in]{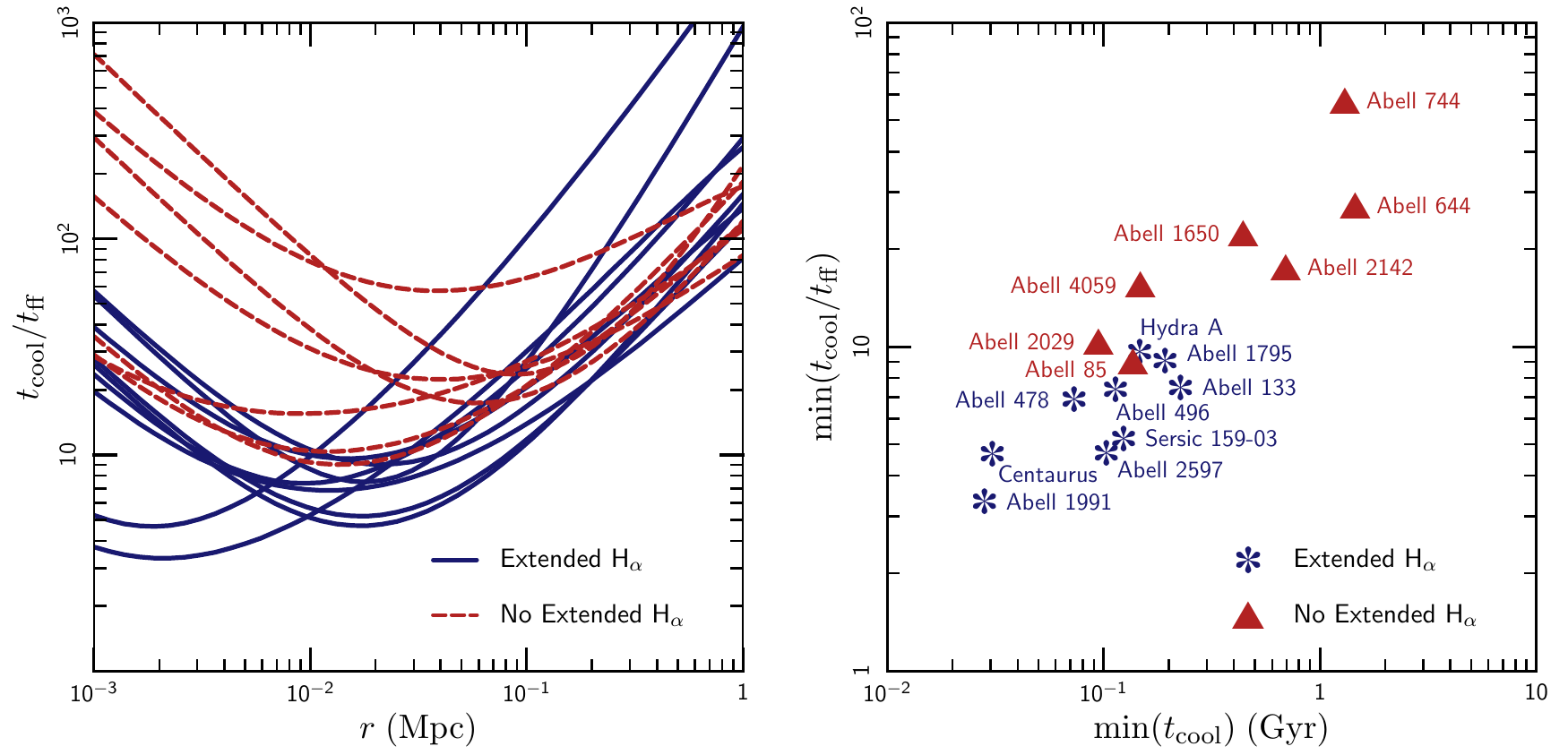}
  \caption{(\textit{Left:}) The timescale ratio $\tsri{cool}{ff}$ as
    a function of radius for clusters in both the \textsc{accept} catalog
    and the \citet{McDonald2010} survey.  Solid blue lines show clusters
    with filaments and dashed red lines show clusters that lack detected
    extended \halpha emission.  Clusters with filaments have systematically
    lower values of $\tsri{cool}{ff}$.  Furthermore, this ratio is smallest
    between $\sim$10--50~kpc, where most filaments are found.
    (\textit{Right:}) The same clusters in the
    $\ts{cool}$--$\tsri{cool}{ff}$ plane.  The coloring is the same as in
    the left panel.  The ratio $\tsri{cool}{ff}$ appears to be a slightly
    better predictor of multi-phase structure than $\ts{cool}$ alone.
    Table~\ref{tab:accept-clusters} lists the clusters plotted in this
    figure.}%
  \label{fig:data}
\end{figure*}
%
\begin{table}
  \caption{Clusters used in~\fig~\ref{fig:data}.}
  \label{tab:accept-clusters}
  \begin{center}
    \hspace*{-1cm}
    \begin{tabular}{cc}
      \toprule
      Extended \halpha & No Extended \halpha \\
      \midrule
      \parbox{\widthof{Abell 1991 .}}{Abell 133}  & \parbox{\widthof{Abell 4059}}{Abell 85}   \\
      \parbox{\widthof{Abell 1991 .}}{Abell 478}  & \parbox{\widthof{Abell 4059}}{Abell 644}  \\
      \parbox{\widthof{Abell 1991 .}}{Abell 496}  & \parbox{\widthof{Abell 4059}}{Abell 744}  \\
      \parbox{\widthof{Abell 1991 .}}{Abell 780}  & \parbox{\widthof{Abell 4059}}{Abell 1650} \\
      \parbox{\widthof{Abell 1991 .}}{Abell 1795} & \parbox{\widthof{Abell 4059}}{Abell 2029} \\
      \parbox{\widthof{Abell 1991 .}}{Abell 1991} & \parbox{\widthof{Abell 4059}}{Abell 2142} \\
      \parbox{\widthof{Abell 1991 .}}{Abell 2597} & \parbox{\widthof{Abell 4059}}{Abell 4059} \\
      Sersic 159-03 & \\
      \parbox{\widthof{Sersic 159-03}}{Centaurus}     & \\
      \bottomrule
    \end{tabular}
  \end{center}

  \medskip
  We use the surveys of \citet{McDonald2010,McDonald2011,McDonald2011b} to
  determine whether a cluster shows multi-phase gas, and we use the data in
  the \textsc{accept} catalog to estimate $\tsri{\ti}{ff}$ for the hot \icm.
  Our label `extended \halpha' signifies that the \halpha emission can be
  resolved and is known to exist outside the \textsc{bcg}; these are the
  Type~I systems from McDonald et\,al.
\end{table}
Observational limits from x-ray spectroscopy \citep{Peterson2006} and from
the shape of the galaxy luminosity function \citep{Benson2003} indicate that
the diffuse plasma in galaxy groups and clusters does not cool as quickly as
it radiates.  These observations imply that some heating process offsets
radiative cooling and that the gas remains in approximate thermal
equilibrium, at least when averaged over length-scales comparable to the
scale-height or time-scales comparable to the cooling time.  The nature of
this ``feedback'' heating is not yet fully understood, although it appears to
involve heating by a central \agn \citep[\eg][]{Birzan2004,McNamara2007}.
More fully understanding the mechanism(s) that regulate the heating to so
closely match cooling remains a major challenge in theories of galaxy
formation.

Though heating strongly suppresses cooling in galaxy groups and clusters,
star formation and multi-phase gas provide clear evidence for cold gas in
many cluster cores.  Observational indicators of this cold gas strongly
correlate with the cooling time of the ambient hot \icm
\citep[\eg][]{Voit2008,Rafferty2008,Cavagnolo2008,Cavagnolo2009}, motivating
a model in which thermal instability in the hot \icm produces much of the
cold gas in cluster cores (as has been suggested many times in the past, \eg
\citealt{Fabian1977,Cowie1980,Nulsen1986,Lowenstein1991}).  Theoretically
studying local thermal instability in the \icm has proven difficult, however,
because of the cooling-flow problem: studies that include both cooling and
gravity typically find that the plasma is \textit{globally} thermally
unstable, and that the entire cluster core collapses monolithically.  This
difficulty has led several authors to conclude that the thermal instability
does not produce multi-phase structure in stably-stratified systems at all.

We avoid the cooling-flow problem in this paper by adopting a
phenomenological heating model that enforces thermal equilibrium when
averaged over large scales~(\S\,\!\ref{subsec:feedback}).  Our heating model
is approximate, over-simplified, and wrong in detail.  However, our results
are insensitive to large temporal and spatial fluctuations about the average
heating~(\S\,\!\ref{subsec:heating-sensitivity}).  Moreover, in
\citetalias{Sharma2011} we obtain similar results using a more physical
feedback heating prescription. We therefore believe that our conclusions
about the saturation of the local thermal instability are reasonably robust.

In the current paradigm in which clusters are approximately in global thermal
equilibrium, heating of the \icm is very likely to depend explicitly on
position in the cluster. In this case, the thermal stability of the plasma is
independent of its convective stability~(\S\,\!\ref{subsec:convection}).
Fundamentally, buoyancy drives convection, while heating and cooling drive
thermal instability. These two processes are formally related only under the
restrictive assumption that heating is a state function of the plasma; more
generally, they are unrelated and the thermal stability of a plasma is
independent of its convective stability \citep[see][]{Balbus1989}.

Nonetheless, it remains true that the competition between buoyancy and
thermal instability determines the net effect of cooling on a stratified
plasma.  We parametrize the relative importance of these effects using the
dimensionless ratio $\tsri{\ti}{ff}$ (the ratio of the thermal instability
growth time $\ts{TI}$ to the local dynamical, or free-fall, time ${t_{\rm
    ff}}$).  When this ratio is small, thermal instability dominates and the
plasma develops significant multi-phase structure; when the ratio
$\tsri{\ti}{ff}$ is large, buoyancy dominates and the plasma remains in a
single, hot phase (\S\,\!\ref{sec:simulation-results}).  This dependence of
the saturation on $\tsri{\ti}{ff}$ is true for {\it both} stably stratified
and neutrally stratified plasmas (\fig~\ref{fig:saturation}), even though the
effect of buoyancy is very different in these two cases.

More quantitatively, we find (\fig~\ref{fig:saturation}) that the saturated
density inhomogeneities produced by the thermal instability approximately
obey the relation
\begin{align}
  \frac{\delta \rho}{\rho} \sim \left(\tsr{\ti}{ff}\right)^{-1} ;
\end{align}
this scaling can be understood analytically by assuming a saturation
amplitude for the thermal instability in which the characteristic fluid
velocities approach $v_{\mr{sat}} \sim H / \ts{cool}$
(\S\,\!\ref{sec:saturation}).  Thus, by assuming that some heating mechanism
prevents cooling catastrophes in clusters, we find that the \icm breaks up
into multiple phases via local thermal instability (with $\delta \rho / \rho
\gtrsim 1$) only if the dimensionless ratio of timescales $\tsri{\ti}{ff}
\lesssim 1$; specifically, there is almost no cold gas at large radii when
$\ts{\ti} \gtrsim \ts{ff}$ (\fig~\ref{fig:saturation}).  This finding is one
of the primary results of our analysis.  We note that the linear growth of
the thermal instability is largely independent of the ratio $\tsri{\ti}{ff}$.
Thus the difference between atmospheres which develop multi-phase gas and
those which do not is fundamentally due to how the \textit{non-linear}
saturation of the thermal instability depends on the atmosphere's properties.

The calculations in \citetalias{Sharma2011} show that the criterion for
multi-phase structure is actually somewhat less stringent in spherical
systems, $\tsri{\ti}{ff} \lesssim 10$.  This difference stems from the fact
that fluid elements are compressed as they move inwards in a spherical
system; this compression enhances the density perturbations and accelerates
the growth of the thermal instability.

Our criterion for multi-phase structure is not sensitive to large variations
about our idealized heating prescription
(\S\,\!\ref{subsec:heating-sensitivity} and
\fig~\ref{fig:heating-fluctuations}).  Furthermore, the multi-phase structure
that develops via thermal instability is largely independent of the magnitude
of the thermal conduction, even on scales much smaller than the Field length
(provided conduction is anisotropic, as is the case in galaxy groups and
clusters; \S\,\!\ref{sec:complex-sims}).  Anisotropic thermal conduction
changes the morphology of the cold gas produced via thermal instability
(blobs $\rightarrow$ filaments), but not the presence or amount of cold gas.
There is thus a very strong, qualitative difference between isotropic and
anisotropic thermal conduction (\fig~\ref{fig:cold-fraction-conduction}),
which cannot be captured by simply multiplying the heat flux by a suppression
factor (as is often done, \eg \citealt{Zakamska2003,Voit2008,Guo2008}).

Our heating prescription imposes global thermal equilibrium and reduces the
accreted mass flux in our model halos relative to cooling-flow values.  This
reduction is not inevitable, however, because thermal instability can produce
cooling-flow-like inflow rates when $\ts{\ti} \lesssim \ts{ff}$.  In a
globally stable system, the thermal instability thus plays an important role
in regulating gas inflow rates.  We study the connections among thermal
instability, mass inflow and feedback more fully in \citetalias{Sharma2011}.

We argue that a locally stable heating mechanism (such as the one proposed in
\citealt{Kunz2011}) is not required to explain the reduced star formation and
cooling rates in clusters.  Instead, global stability arising from
approximate thermal equilibrium, together with the physics of local thermal
instability in stratified plasmas, is sufficient to reproduce the low net
cooling rates in clusters.  Moreover, the correlation of \halpha filaments
and star formation in clusters with the cooling time in the hot \icm strongly
suggests that the plasma is in fact locally thermally unstable.  We have
shown that the suppression of accretion rates is not sensitive to thermal
conduction or to significant variations about our specific feedback
prescription~(\S\,\!\ref{subsec:heating-sensitivity} and
\fig~\ref{fig:heating-fluctuations}), and we explore this further in
\citetalias{Sharma2011}.

We now compare our model predictions with observational results.  The thermal
instability time $\ts{\ti} \equiv \gr{\ti}^{-1}$
(eq.~\ref{eq:field-growth-rate}) depends on the unknown way in which feedback
energy is thermalized (eq.~\ref{eq:heating-fn}) and cannot be directly
inferred from observations.  We therefore use the cooling time when comparing
our results with observations.  These two timescales differ by an unknown
factor of order unity.  The ratio of timescales $\tsri{cool}{ff}$ can be
reexpressed in more familiar terms using
\begin{align}
  \tsr{cool}{ff} \sim 3
  \frac{\left(K / 10~\unit{keV~cm^2}\right)^{3/2}}%
  {T_7^{1/2} \, \Lambda_{-23} \, \left(\ts{ff} / 30~\unit{Myr}\right)},
\end{align}
where $K = \kb T/n^{2/3}$ is a measure of the plasma entropy, $T_7$ is the
temperature in units of $10^7\,\unit{K}$ and $\Lambda_{-23}$ is the cooling
function (eq.~\ref{eq:raw-cooling-fn}) in units of
$10^{-23}\unit{~erg~cm^{3}~s^{-1}}$.  Thus, the plasma in clusters and
galactic halos should show extended multi-phase structure wherever
\begin{align}
  K \lesssim \left(20~\unit{keV~cm^2}\right)
  \left[ T_7^{1/2} \, \Lambda_{-23} \,
    \left(\frac{\ts{ff}}{30~\unit{Myr}}\right) \right]^{2/3}
  ,\label{eq:entropy-criterion}
\end{align}
where we have used the threshold from \citetalias{Sharma2011} for multi-phase
structure in spherical systems: $\tsri{\ti}{ff} \sim 10$.  This criterion is
consistent with observations that clusters with central entropies below
$30~\unit{keV~cm^2}$ preferentially show signs of cold gas such as star
formation and \halpha emission \citep{Voit2008,Cavagnolo2008,Cavagnolo2009}.
Note, however, the relatively strong dependence of this criterion on
$\ts{ff}$ and on $\Lambda$; this is because the entropy $K$ is not the
fundamental parameter governing the saturation of the thermal instability in
a stratified system.

The \halpha survey conducted by \citet{McDonald2010,McDonald2011} permits
another test of our criterion.  McDonald et\,al\@. provide lists of groups
and clusters with and without extended \halpha emission.  We test our
criterion by estimating the time-scale ratio $\tsri{cool}{ff}$ for these
systems using data from the \textsc{accept} catalog \citep{Cavagnolo2009}.
We fit the entropy profiles of clusters in the \textsc{accept} catalog using
$K(r) = K_0 + K_1 (r/100\unit{kpc})^a$ and we fit the pressure profiles
$P(r)$ using the form provided in \citet{Arnaud2010}.  Of the groups and
clusters in both the \halpha surveys and in \textsc{accept}, sixteen give
reasonable fits (listed in
Table~\ref{tab:accept-clusters}).\footnote{Unfortunately, we were unable to
  fit several clusters with well-known filament systems, including Perseus,
  Abell~2052 and M87.  The pressure gradients in Perseus and Abell~2052 are
  positive at some radii and cannot be fit by the universal profile.
  Similarly, the pressure profile for M87 deviates from the broken power-law
  universal profile.} From our fits to $K(r)$ and $P(r)$, we calculate $n(r)$
and $T(r)$ and estimate $\ts{cool}(r)$ using the fit to the cooling function
provided by \citet{Tozzi2001} with $1/3$ solar metallicity.  We estimate
$g(r)$ from the pressure and density profiles by assuming spherical symmetry
and hydrostatic equilibrium; from this we calculate $\ts{ff}(r)$.

The left panel of \fig~\ref{fig:data} shows $\tsri{cool}{ff}$ as a function
of radius for these sixteen groups and clusters.  As predicted by our
analysis, the clusters with short cooling times $\tsri{cool}{ff} \lesssim 10$
show extended filaments, while the clusters with long cooling times
$\tsri{cool}{ff} \gtrsim 10$ do not.  Additionally, most of the filaments are
found at radii around 10--50~kpc, where the ratio $\tsri{\ti}{ff}$ is the
smallest.  Although this evidence is not conclusive, these data support our
hypothesis that the filaments condense from the \icm due to the local thermal
instability.

We emphasize that both cooling and gravity influence the development of the
thermal instability in the \icm.  A short cooling time (or low $K$) is not
sufficient for the formation of filaments; rather, the ratio $\tsri{\ti}{ff}$
is the relevant parameter.  The right panel of \fig~\ref{fig:data} shows
clusters in the ($\ts{cool}$)--($\tsri{cool}{ff}$) plane.  More data are
needed to conclusively test our model, but these results are consistent with
our interpretation that the ratio of timescales $\tsri{cool}{ff}$ is a better
predictor of multi-phase gas in hot halos than $\ts{cool}$ alone.

These simple comparisons support a model in which local thermal instability
produces at least some of the \halpha filaments seen in clusters.  This does
not, however, imply that thermal instability alone can explain all of the
observed properties of multi-phase gas in clusters and/or galaxies.  On the
contrary, processes such as conductive condensation of hot gas to cool gas
(``non-radiative cooling;'' \citealt{Fabian2002,Soker2004}) and the inflow of
cold gas through the virial radius \citep{Keres2009} may also be
important (in higher and lower mass halos, respectively).  Furthermore, other
processes in the \icm such as merger shocks, galaxy wakes and buoyant radio
bubbles may influence the evolution of the filaments.

Accretion of the cold gas formed via thermal instability likely plays an
important role in the evolution of brightest cluster galaxies and their
central \agn \citep{Pizzolato2005,Pizzolato2010}.  In addition, the
high-velocity clouds surrounding the Milky Way may also be manifestations of
the thermal instability \citep{Maller2004,Sommer2006,Kaufmann2006,Peek2008};
this process could provide an important source of unenriched gas to maintain
metallicity gradients \citep{Jones2010} and continued star formation
\citep{Bauermeister2010} in the Milky Way and other galaxies.

\acknowledgments
\noindent We are grateful to Mark Voit and Megan Donahue for interesting and
helpful conversations as we completed this work, and to Steve Balbus for
clarifying our discussion of the previous literature.  We thank the anonymous
referee for helping to clarify our discussion, particularly in
section~\ref{sec:linear-results}.  Mike McDonald helpfully suggested several
clusters we left out from \fig~\ref{fig:data} in an earlier version of this
paper.  We are also thankful for the hospitality of the Kavli Institute for
Theoretical Physics (KITP) at UC Santa Barbara, where we performed some of
this work.  KITP is supported in part by NSF grant number PHY$05$-$51164$.
Support for P.~S. was provided by NASA through Chandra Postdoctoral Fellowship
grant PF$8$-$90054$ awarded by the Chandra X-Ray Center, which is operated by
the Smithsonian Astrophysical Observatory for NASA under contract
NAS$8$-$03060$.  M.~M., I.~P. and E.~Q. were supported in part by NASA Grant
NNX$10$AC$95$G, NSF-DOE Grant PHY-$0812811$, and by the David and Lucile
Packard Foundation.  We performed many of the computations for this paper on
the \textit{Henyey} cluster at UC Berkeley, supported by NSF AST Grant
$0905801$; additional computing time was provided by the National Science
Foundation through the Teragrid resources located at the National Center for
Atmospheric Research under grant number TG-AST$090038$.  We made our figures
using the open-source program \textsc{Tioga}.  This research has made use of
NASA's Astrophysics Data System. \\

\bibliography{thermal_instability}

\begin{thebibliography}{75}
\expandafter\ifx\csname natexlab\endcsname\relax\def\natexlab#1{#1}\fi

\bibitem[{{Arnaud} {et~al.}(2010){Arnaud}, {Pratt}, {Piffaretti},
  {B{\"o}hringer}, {Croston}, \& {Pointecouteau}}]{Arnaud2010}
{Arnaud}, M., {Pratt}, G.~W., {Piffaretti}, R., {et~al.} 2010, \aap, 517, A92

\bibitem[{{Balbus}(1986)}]{Balbus1986}
{Balbus}, S.~A. 1986, \apjl, 303, L79

\bibitem[{{Balbus}(1988)}]{Balbus1988}
{Balbus}, S.~A. 1988, \apj, 328, 395

\bibitem[{{Balbus}(2000)}]{Balbus2000}
{Balbus}, S.~A. 2000, \apj, 534, 420

\bibitem[{{Balbus}(2001)}]{Balbus2001}
{Balbus}, S.~A. 2001, \apj, 562, 909

\bibitem[{{Balbus} \& {Reynolds}(2010)}]{Balbus2010}
{Balbus}, S.~A. \& {Reynolds}, C.~S. 2010, \apjl, 720, L97

\bibitem[{{Balbus} \& {Soker}(1989)}]{Balbus1989}
{Balbus}, S.~A. \& {Soker}, N. 1989, \apj, 341, 611

\bibitem[{{Bauermeister} {et~al.}(2010){Bauermeister}, {Blitz}, \&
  {Ma}}]{Bauermeister2010}
{Bauermeister}, A., {Blitz}, L., \& {Ma}, C. 2010, \apj, 717, 323

\bibitem[{{Benson} {et~al.}(2003){Benson}, {Bower}, {Frenk}, {Lacey}, {Baugh},
  \& {Cole}}]{Benson2003}
{Benson}, A.~J., {Bower}, R.~G., {Frenk}, C.~S., {et~al.} 2003, \apj, 599, 38

\bibitem[{{Binney} {et~al.}(2009){Binney}, {Nipoti}, \&
  {Fraternali}}]{Binney2009}
{Binney}, J., {Nipoti}, C., \& {Fraternali}, F. 2009, \mnras, 397, 1804

\bibitem[{{B{\^i}rzan} {et~al.}(2004){B{\^i}rzan}, {Rafferty}, {McNamara},
  {Wise}, \& {Nulsen}}]{Birzan2004}
{B{\^i}rzan}, L., {Rafferty}, D.~A., {McNamara}, B.~R., {Wise}, M.~W., \&
  {Nulsen}, P.~E.~J. 2004, \apj, 607, 800

\bibitem[{{Braginskii}(1965)}]{Braginskii1965}
{Braginskii}, S.~I. 1965, Reviews of Plasma Physics, 1, 205

\bibitem[{{Cavagnolo} {et~al.}(2008){Cavagnolo}, {Donahue}, {Voit}, \&
  {Sun}}]{Cavagnolo2008}
{Cavagnolo}, K.~W., {Donahue}, M., {Voit}, G.~M., \& {Sun}, M. 2008, \apjl,
  683, L107

\bibitem[{{Cavagnolo} {et~al.}(2009){Cavagnolo}, {Donahue}, {Voit}, \&
  {Sun}}]{Cavagnolo2009}
{Cavagnolo}, K.~W., {Donahue}, M., {Voit}, G.~M., \& {Sun}, M. 2009, \apjs,
  182, 12

\bibitem[{{Ciotti} \& {Ostriker}(2001)}]{Ciotti2001}
{Ciotti}, L. \& {Ostriker}, J.~P. 2001, \apj, 551, 131

\bibitem[{{Cole} {et~al.}(2001){Cole}, {Norberg}, {Baugh}, {Frenk},
  {Bland-Hawthorn}, {Bridges}, {Cannon}, {Colless}, {Collins}, {Couch},
  {Cross}, {Dalton}, {De Propris}, {Driver}, {Efstathiou}, {Ellis},
  {Glazebrook}, {Jackson}, {Lahav}, {Lewis}, {Lumsden}, {Maddox}, {Madgwick},
  {Peacock}, {Peterson}, {Sutherland}, \& {Taylor}}]{Cole2001}
{Cole}, S., {Norberg}, P., {Baugh}, C.~M., {et~al.} 2001, \mnras, 326, 255

\bibitem[{{Cowie} {et~al.}(1980){Cowie}, {Fabian}, \& {Nulsen}}]{Cowie1980}
{Cowie}, L.~L., {Fabian}, A.~C., \& {Nulsen}, P.~E.~J. 1980, \mnras, 191, 399

\bibitem[{{Davis} {et~al.}(2010){Davis}, {Stone}, \& {Pessah}}]{Davis2010}
{Davis}, S.~W., {Stone}, J.~M., \& {Pessah}, M.~E. 2010, \apj, 713, 52

\bibitem[{{Defouw}(1970)}]{Defouw1970}
{Defouw}, R.~J. 1970, \apj, 160, 659

\bibitem[{{Fabian}(1994)}]{Fabian1994}
{Fabian}, A.~C. 1994, \araa, 32, 277

\bibitem[{{Fabian} {et~al.}(2002){Fabian}, {Allen}, {Crawford}, {Johnstone},
  {Morris}, {Sanders}, \& {Schmidt}}]{Fabian2002}
{Fabian}, A.~C., {Allen}, S.~W., {Crawford}, C.~S., {et~al.} 2002, \mnras, 332,
  L50

\bibitem[{{Fabian} {et~al.}(2008){Fabian}, {Johnstone}, {Sanders}, {Conselice},
  {Crawford}, {Gallagher}, \& {Zweibel}}]{Fabian2008}
{Fabian}, A.~C., {Johnstone}, R.~M., {Sanders}, J.~S., {et~al.} 2008, \nat,
  454, 968

\bibitem[{{Fabian} \& {Nulsen}(1977)}]{Fabian1977}
{Fabian}, A.~C. \& {Nulsen}, P.~E.~J. 1977, \mnras, 180, 479

\bibitem[{{Fabian} {et~al.}(2003){Fabian}, {Sanders}, {Crawford}, {Conselice},
  {Gallagher}, \& {Wyse}}]{Fabian2003}
{Fabian}, A.~C., {Sanders}, J.~S., {Crawford}, C.~S., {et~al.} 2003, \mnras,
  344, L48

\bibitem[{{Ferland} {et~al.}(2009){Ferland}, {Fabian}, {Hatch}, {Johnstone},
  {Porter}, {van Hoof}, \& {Williams}}]{Ferland2009}
{Ferland}, G.~J., {Fabian}, A.~C., {Hatch}, N.~A., {et~al.} 2009, \mnras, 392,
  1475

\bibitem[{{Field}(1965)}]{Field1965}
{Field}, G.~B. 1965, \apj, 142, 531

\bibitem[{{Fromang} \& {Papaloizou}(2007)}]{Fromang2007}
{Fromang}, S. \& {Papaloizou}, J. 2007, \aap, 476, 1113

\bibitem[{{Guo} \& {Oh}(2008)}]{Guo2008}
{Guo}, F. \& {Oh}, S.~P. 2008, \mnras, 384, 251

\bibitem[{{Hattori} \& {Habe}(1990)}]{Hattori1990}
{Hattori}, M. \& {Habe}, A. 1990, \mnras, 242, 399

\bibitem[{{Heckman} {et~al.}(1989){Heckman}, {Baum}, {van Breugel}, \&
  {McCarthy}}]{Heckman1989}
{Heckman}, T.~M., {Baum}, S.~A., {van Breugel}, W.~J.~M., \& {McCarthy}, P.
  1989, \apj, 338, 48

\bibitem[{{Holtzman} {et~al.}(1992){Holtzman}, {Faber}, {Shaya}, {Lauer},
  {Groth}, {Hunter}, {Baum}, {Ewald}, {Hester}, {Light}, {Lynds}, {O'Neil}, \&
  {Westphal}}]{Holtzman1992}
{Holtzman}, J.~A., {Faber}, S.~M., {Shaya}, E.~J., {et~al.} 1992, \aj, 103, 691

\bibitem[{{Hu} {et~al.}(1985){Hu}, {Cowie}, \& {Wang}}]{Hu1985}
{Hu}, E.~M., {Cowie}, L.~L., \& {Wang}, Z. 1985, \apjs, 59, 447

\bibitem[{{Jones} {et~al.}(2010){Jones}, {Ellis}, {Jullo}, \&
  {Richard}}]{Jones2010}
{Jones}, T., {Ellis}, R., {Jullo}, E., \& {Richard}, J. 2010, \apjl, 725, L176

\bibitem[{{Joung} {et~al.}(2011){Joung}, {Bryan}, \& {Putman}}]{Joung2011}
{Joung}, M.~R., {Bryan}, G.~L., \& {Putman}, M.~E. 2011, ArXiv e-prints

\bibitem[{{Kaufmann} {et~al.}(2006){Kaufmann}, {Mayer}, {Wadsley}, {Stadel}, \&
  {Moore}}]{Kaufmann2006}
{Kaufmann}, T., {Mayer}, L., {Wadsley}, J., {Stadel}, J., \& {Moore}, B. 2006,
  \mnras, 370, 1612

\bibitem[{{Kere{\v s}} \& {Hernquist}(2009)}]{Keres2009}
{Kere{\v s}}, D. \& {Hernquist}, L. 2009, \apjl, 700, L1

\bibitem[{{Kochanek} {et~al.}(2001){Kochanek}, {Pahre}, {Falco}, {Huchra},
  {Mader}, {Jarrett}, {Chester}, {Cutri}, \& {Schneider}}]{Kochanek2001}
{Kochanek}, C.~S., {Pahre}, M.~A., {Falco}, E.~E., {et~al.} 2001, \apj, 560,
  566

\bibitem[{{Kunz} {et~al.}(2011){Kunz}, {Schekochihin}, {Cowley}, {Binney}, \&
  {Sanders}}]{Kunz2011}
{Kunz}, M.~W., {Schekochihin}, A.~A., {Cowley}, S.~C., {Binney}, J.~J., \&
  {Sanders}, J.~S. 2011, \mnras, 410, 2446

\bibitem[{{Loewenstein} {et~al.}(1991){Loewenstein}, {Zweibel}, \&
  {Begelman}}]{Lowenstein1991}
{Loewenstein}, M., {Zweibel}, E.~G., \& {Begelman}, M.~C. 1991, \apj, 377, 392

\bibitem[{{Lynds} \& {Sandage}(1963)}]{Lynds1963}
{Lynds}, C.~R. \& {Sandage}, A.~R. 1963, \apj, 137, 1005

\bibitem[{{Lynds}(1970)}]{Lynds1970}
{Lynds}, R. 1970, \apjl, 159, L151

\bibitem[{{Malagoli} {et~al.}(1987){Malagoli}, {Rosner}, \&
  {Bodo}}]{Malagoli1987}
{Malagoli}, A., {Rosner}, R., \& {Bodo}, G. 1987, \apj, 319, 632

\bibitem[{{Malagoli} {et~al.}(1990){Malagoli}, {Rosner}, \&
  {Fryxell}}]{Malagoli1990}
{Malagoli}, A., {Rosner}, R., \& {Fryxell}, B. 1990, \mnras, 247, 367

\bibitem[{{Maller} \& {Bullock}(2004)}]{Maller2004}
{Maller}, A.~H. \& {Bullock}, J.~S. 2004, \mnras, 355, 694

\bibitem[{{McCarthy} {et~al.}(2004){McCarthy}, {Balogh}, {Babul}, {Poole}, \&
  {Horner}}]{McCarthy2004}
{McCarthy}, I.~G., {Balogh}, M.~L., {Babul}, A., {Poole}, G.~B., \& {Horner},
  D.~J. 2004, \apj, 613, 811

\bibitem[{{McCourt} {et~al.}(2011){McCourt}, {Parrish}, {Sharma}, \&
  {Quataert}}]{McCourt2011}
{McCourt}, M., {Parrish}, I.~J., {Sharma}, P., \& {Quataert}, E. 2011, \mnras,
  413, 1295

\bibitem[{{McDonald} {et~al.}(2011{\natexlab{a}}){McDonald}, {Veilleux}, \&
  {Mushotzky}}]{McDonald2011}
{McDonald}, M., {Veilleux}, S., \& {Mushotzky}, R. 2011{\natexlab{a}}, \apj,
  731, 33

\bibitem[{{McDonald} {et~al.}(2010){McDonald}, {Veilleux}, {Rupke}, \&
  {Mushotzky}}]{McDonald2010}
{McDonald}, M., {Veilleux}, S., {Rupke}, D.~S.~N., \& {Mushotzky}, R. 2010,
  \apj, 721, 1262

\bibitem[{{McDonald} {et~al.}(2011{\natexlab{b}}){McDonald}, {Veilleux},
  {Rupke}, {Mushotzky}, \& {Reynolds}}]{McDonald2011b}
{McDonald}, M., {Veilleux}, S., {Rupke}, D.~S.~N., {Mushotzky}, R., \&
  {Reynolds}, C. 2011{\natexlab{b}}, \apj, 734, 95

\bibitem[{{McNamara} \& {Nulsen}(2007)}]{McNamara2007}
{McNamara}, B.~R. \& {Nulsen}, P.~E.~J. 2007, \araa, 45, 117

\bibitem[{{Nulsen}(1986)}]{Nulsen1986}
{Nulsen}, P.~E.~J. 1986, \mnras, 221, 377

\bibitem[{{O'Dea} {et~al.}(2010){O'Dea}, {Quillen}, {O'Dea}, {Tremblay},
  {Snios}, {Baum}, {Christiansen}, {Noel-Storr}, {Edge}, {Donahue}, \&
  {Voit}}]{ODea2010}
{O'Dea}, K.~P., {Quillen}, A.~C., {O'Dea}, C.~P., {et~al.} 2010, \apj, 719,
  1619

\bibitem[{{Oh} \& {Benson}(2003)}]{Oh2003}
{Oh}, S.~P. \& {Benson}, A.~J. 2003, \mnras, 342, 664

\bibitem[{{Parrish} \& {Stone}(2005)}]{Parrish2005}
{Parrish}, I.~J. \& {Stone}, J.~M. 2005, \apj, 633, 334

\bibitem[{{Peek} {et~al.}(2008){Peek}, {Putman}, \& {Sommer-Larsen}}]{Peek2008}
{Peek}, J.~E.~G., {Putman}, M.~E., \& {Sommer-Larsen}, J. 2008, \apj, 674, 227

\bibitem[{{Peterson} \& {Fabian}(2006)}]{Peterson2006}
{Peterson}, J.~R. \& {Fabian}, A.~C. 2006, \physrep, 427, 1

\bibitem[{{Pizzolato} \& {Soker}(2005)}]{Pizzolato2005}
{Pizzolato}, F. \& {Soker}, N. 2005, \apj, 632, 821

\bibitem[{{Pizzolato} \& {Soker}(2010)}]{Pizzolato2010}
{Pizzolato}, F. \& {Soker}, N. 2010, \mnras, 408, 961

\bibitem[{{Quataert}(2008)}]{Quataert2008}
{Quataert}, E. 2008, \apj, 673, 758

\bibitem[{{Rafferty} {et~al.}(2008){Rafferty}, {McNamara}, \&
  {Nulsen}}]{Rafferty2008}
{Rafferty}, D.~A., {McNamara}, B.~R., \& {Nulsen}, P.~E.~J. 2008, \apj, 687,
  899

\bibitem[{{Rees} \& {Ostriker}(1977)}]{Rees1977}
{Rees}, M.~J. \& {Ostriker}, J.~P. 1977, \mnras, 179, 541

\bibitem[{{Saro} {et~al.}(2006){Saro}, {Borgani}, {Tornatore}, {Dolag},
  {Murante}, {Biviano}, {Calura}, \& {Charlot}}]{Saro2006}
{Saro}, A., {Borgani}, S., {Tornatore}, L., {et~al.} 2006, \mnras, 373, 397

\bibitem[{{Sharma} \& {Hammett}(2007)}]{Sharma2007}
{Sharma}, P. \& {Hammett}, G.~W. 2007, Journal of Computational Physics, 227,
  123

\bibitem[{{Sharma} {et~al.}(2012){Sharma}, {McCourt}, {Quataert}, \&
  {Parrish}}]{Sharma2011}
{Sharma}, P., {McCourt}, M., {Quataert}, E., \& {Parrish}, I.~J. 2012, \mnras,
  420, 3174

\bibitem[{{Sharma} {et~al.}(2010){Sharma}, {Parrish}, \&
  {Quataert}}]{Sharma2010}
{Sharma}, P., {Parrish}, I.~J., \& {Quataert}, E. 2010, \apj, 720, 652

\bibitem[{{Sijacki} \& {Springel}(2006)}]{Sijacki2006}
{Sijacki}, D. \& {Springel}, V. 2006, \mnras, 366, 397

\bibitem[{{Silk}(1977)}]{Silk1977}
{Silk}, J. 1977, \apj, 211, 638

\bibitem[{{Soker}(2006)}]{Soker2006}
{Soker}, N. 2006, \na, 12, 38

\bibitem[{{Soker} {et~al.}(2004){Soker}, {Blanton}, \& {Sarazin}}]{Soker2004}
{Soker}, N., {Blanton}, E.~L., \& {Sarazin}, C.~L. 2004, \aap, 422, 445

\bibitem[{{Sommer-Larsen}(2006)}]{Sommer2006}
{Sommer-Larsen}, J. 2006, \apjl, 644, L1

\bibitem[{{Spitzer}(1962)}]{Spitzer1962}
{Spitzer}, L. 1962, {Physics of Fully Ionized Gases} ({New York}: Interscience
  Publishers)

\bibitem[{{Tozzi} \& {Norman}(2001)}]{Tozzi2001}
{Tozzi}, P. \& {Norman}, C. 2001, \apj, 546, 63

\bibitem[{{Voit} {et~al.}(2008){Voit}, {Cavagnolo}, {Donahue}, {Rafferty},
  {McNamara}, \& {Nulsen}}]{Voit2008}
{Voit}, G.~M., {Cavagnolo}, K.~W., {Donahue}, M., {et~al.} 2008, \apjl, 681, L5

\bibitem[{{White} \& {Rees}(1978)}]{White1978}
{White}, S.~D.~M. \& {Rees}, M.~J. 1978, \mnras, 183, 341

\bibitem[{{Zakamska} \& {Narayan}(2003)}]{Zakamska2003}
{Zakamska}, N.~L. \& {Narayan}, R. 2003, \apj, 582, 162

\end{thebibliography}

\end{document}